\documentclass[twocolumn, asmath, amssymb, floatfix, superscriptaddress]{revtex4-2}
\usepackage[dvips]{graphics}
\usepackage[justification=raggedright]{caption}
\usepackage{subcaption}
\usepackage[font=footnotesize,figurewithin=none]{caption}
\usepackage{adjustbox}
\usepackage[labelfont=bf]{caption}
\usepackage[labelformat=parens, labelfont=bf]{subcaption}
\usepackage{color}
\usepackage{float}
\usepackage{natbib}
\usepackage[dvipsnames]{xcolor}
\setcitestyle{numbers}
\setcitestyle{square}
\definecolor{dred}{rgb}{0,0,0.6}
\definecolor{NavyBlue}{rgb}{0, 0, 128}
\usepackage{graphicx}
\usepackage{bm}
\usepackage{amssymb}
\usepackage{amsmath}
\usepackage{mathtools}
\usepackage{dcolumn}   
\usepackage[colorlinks, linkcolor=NavyBlue,citecolor=NavyBlue,urlcolor=NavyBlue]{hyperref}
\usepackage[all]{hypcap} 
\usepackage{physics}
\usepackage{enumerate}
\usepackage{braket}        
\usepackage{overpic}
\usepackage{amsfonts}
\usepackage{dsfont}
\usepackage[mathscr]{eucal}
\usepackage{hyperref}
\usepackage{setspace}

\newcommand{\dg}{\dagger}
\newcommand{\mb}{\mathbf}

\newcommand{\prbsubref}[2]{\hyperref[#1]{\ref*{#1}{(\subref*{#2})}}}

\begin{document}
\title{Studying magnon band topology through low-energy magnon excitations: role of anisotropic Dzyaloshinskii-Moriya interaction }

\author{Shreya Debnath and Saurabh Basu\\ \textit{Department of Physics, Indian Institute of Technology Guwahati, Guwahati-781039, Assam, India}}

\begin{abstract}
In this work, we study topological properties of magnons via creating spin excitations in both ferromagnets and antiferromagnets in presence of an external magnetic field on a two-dimensional square lattice. It is known that Dzyaloshinskii-Moriya interaction (DMI) plays an important role in coupling between different particle (spin excitation) sectors, here we consider an anisotropic DMI and ascertain the role of the anisotropy parameter in inducing topological phase transitions. While the scenario, for dealing with ferromagnets, albeit with isotropic DMI is established in literature, we have developed the formalism for studying magnon band topology for the antiferromagnetic case. The calculations for the ferromagnetic case are included to facilitate a comparison between the two magnetically ordered systems. Owing to the presence of a two-sublattice structure of an antiferromagnet, a larger number of magnon bands participate in deciding upon the topological properties. However, in both the cases, an extended trivial region is observed even with the DMI to be non-zero, which is surprising since the DMI is the origin of the finite Berry curvature in presence of external magnetic field. Furthermore, in an antiferromagnet, a smaller anisotropy is capable of inducing a gap-closing transition from a topological to a trivial phase compared to that for a ferromagnet. The nature of the phases in both the cases and the phase transitions therein are characterized by the band structure, presence (or absence) of the chiral edge modes observed in a semi-infinite nano-ribbon geometry, computation of the thermal Hall effect, etc. Moreover, the strength of the magnetic field is found to play a decisive role in controlling the critical point that demarcates various topological phases.
\end{abstract}

\maketitle


\section{INTRODUCTION}

Magnetic materials offer an ideal platform to study nontrivial band topology in different systems, thereby giving a boost to explain their spectral and transport properties.~In the emergent era of spintronic devices~\cite{spintronics1_2018, spintronics2_2022,spintronics4_2023,spintronics5_2020,spintronics6_2019}, skyrmions~\cite{skyrmion3_2022,skyrmion4_2021,skyrmion5_2023}, spin wave excitations, such as magnons~\cite{magnon1_2022,magnon2_2020,magnon3_2021,magnon4_2016,magnon5_2019,magnon6_2021, zhuo2023topological},  etc, have recently taken centerstage as candidates for spin-based information technology~\cite{magnon-skyrmion,magnon1,skyrmion1,skyrmion2}.~It uses charge-neutral magnetic excitations that are unaffected by Coulomb interactions.~In order to delve into the study of spin wave excitations in quantum magnets~\cite{spinwave-qmagnet}, we need to develop a framework for our purpose that encapsulates the key elements of magnon-band topology.~Apart from a nearest neighbour (and may be a next nearest neighbour) Heisenberg-type exchange interaction, an asymmetric Dzyaloshinskii-Moriya interaction (DMI)~\cite{dzyaloshinsky1958, moriya1960, yang2023first} plays an important role in magnetic systems. DMI promotes topological spin excitations and is responsible for the observation of thermal Hall effect (THE)~\cite{thermalHallantiferro, sugii2017thermal, hirokane2019phononic, li2020phonon, thermalhall2010}, which is analogous, yet distinct from the topological (quantized) Hall effect observed in fermionic systems.~DMI is particularly strong in systems that lack inversion symmetry and favour chiral configurations~\cite{DMIfavourcanting2001}, such as different types of skyrmions, anti-skyrmions etc. The DMI term can be written as~\cite{moriya1960},
\begin{equation}
    H_{\text{\tiny{DMI}}}=\sum_{\braket{p,q}}\mathbf{D}_{pq}\cdot(\mathbf{S}_p \times \mathbf{S}_q),\label{DMI}
\end{equation}
where $\mathbf{D}_{pq}$ denotes the Dzyaloshinskii-Moriya (DM) vector and $\mathbf{S}_p$ and $\mathbf{S}_q$ refer to the spin angular momenta at sites $p$ and $q$.

In the presence of the DMI, topological magnons have been studied in various types of magnetically ordered systems~\cite{moulsdale2019unconventional, zhuo2021topological, li2021magnonic, czajka2023planar}, including ferromagnets, antiferromagnets, and some frustrated magnets~\cite{zhang2021topological}. An alternative approach to studying magnon band topology involves examining spin excitations in different particle sectors. These spin excitations are known as bosonic quasiparticles, characterized by particle creation operators $a_i^\dagger$ and annihilation operators $a_i$. In a perfectly ordered magnetic system, a single spin-flip is identified as the lowest-energy single-magnon (one-magnon) state, corresponding to a single-particle sector,  whereas two such spin-flips yield a two-magnon state in the two-particle sector. The energy of a two-magnon state generally consists of two one-magnon energies, that form a two-magnon continuum for a finite lattice. In cases where two spin-flips occur next to each other, a bound state can emerge with an energy lower than that of two one-magnon energies. If the binding energy becomes sufficiently strong for the two-magnon bound states to be situated near the one-magnon energy spectrum and well below the two-magnon continuum, these two-magnon bound states behave as quasiparticles, leading to nontrivial band topology under certain circumstances. 

In this paper, we primarily focus on the topological properties that arise due to the hybridization between different particle sectors, especially between one-magnon and two-magnon bound states, which define low-energy spin excitations.
In order to allow this hybridization, the system must break $U(1)$ symmetry, which is related to the particle number conservation in the system. In the following section, we shall discuss how DMI plays an important role in breaking this $U(1)$ symmetry.

Let us discuss the theoretical and experimental scenarios that have motivated us to study how DMI affects the magnon band topology. In a two-dimensional (2D) system, DMI is usually considered as an isotropic quantity, namely $D_x=D_y$, where $x$ and $y$ refer to the two spatial directions. In this case, the theoretical foundations for analyses of magnon band topology via hybridization between different particle sectors have been introduced by the authors of Ref.~\cite{magnonboundpairs2023}. In addition to the theoretical studies, experimental observations of such hybridization have been reported. One-magnon and two-magnon excitations have been investigated in a one-dimensional (1D) antiferromagnet $(\text{CD}_3)_4\text{NMnCl}_3$~\cite{heilmann1981one} and a quasi-1D antiferromagnet $\text{CsVCl}_3$~\cite{inami1997observation}. A recent study on the spin-1 antiferromagnet $\text{FeI}_2$~\cite{bai2021hybridized} revealed hybridization between a single-magnon and a single-ion bound state. Whereas, in a spin-$\frac{1}{2}$ quantum magnet $\text{SrCu}_2(\text{BO}_3)_2$~\cite{mcclarty2017topological}, a nontrivial single-triplon spectrum has been studied.

An anisotropic DMI in such scenarios may yield observable consequences on the band structure and hence may have important ramifications on the topological properties. Furthermore, the presence of anisotropic DMI has been identified in various recent studies. ~In an out-of-plane magnetized Au/Co/W(110) heterostructures, the DMI strength is found to be twice (or even larger) along the [$\bar{1}10$] direction than that along the [001] direction~\cite{S3}. Different from the isotropic DMI ($D_x=D_y$) case, an anisotropic DMI with same magnitude, but mutually opposite signs, that is, $D_x=-D_y$ may exist in 2D ternary compounds, such as, $\text{MCuX}_2$ (M: transition metal, X: group VIA)~\cite{S4,S5}. Along with these material studies, the presence of an anisotropic DMI has also been studied theoretically in bulk systems with $D_{2d}$ and $S_4$ symmetries~\cite{hoffmann}. A recent study shows that in FM/NM/FM (FM: ferromagnetic, NM: non-magnetic) structures, where two separate FM layers can couple to each other ferromagnetically or antiferromagnetically, the interlayer DMI can be tuned by the NM metal layer thickness~\cite{yun2023anisotropic}. Further, spin-1/2 triangular-lattice Heisenberg antiferromagnet $\text{Ba}_3\text{CoSb}_2\text{O}_9$~\cite{ito2017structure} can be interesting to us, where owing to the highly symmetric crystal structure, DMI is absent between neighboring spins but may possess longer range DMI. Also, recent studies show that DMI can be induced and controlled via growing the sample on a substrate, which is responsible for symmetry breaking at the surface and could be a source of strong spin-orbit coupling.

\begin{figure}[t]
     \begin{subfigure}[!h]{0.49\columnwidth}
         \includegraphics[width=\columnwidth]{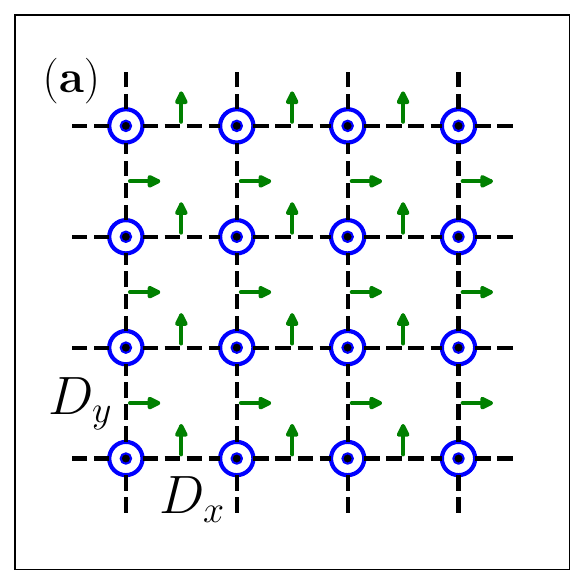}
         \captionlistentry{}
         \label{ferromagnet}
     \end{subfigure}
     \begin{subfigure}[!h]{0.49\columnwidth}
         \includegraphics[width=\columnwidth]{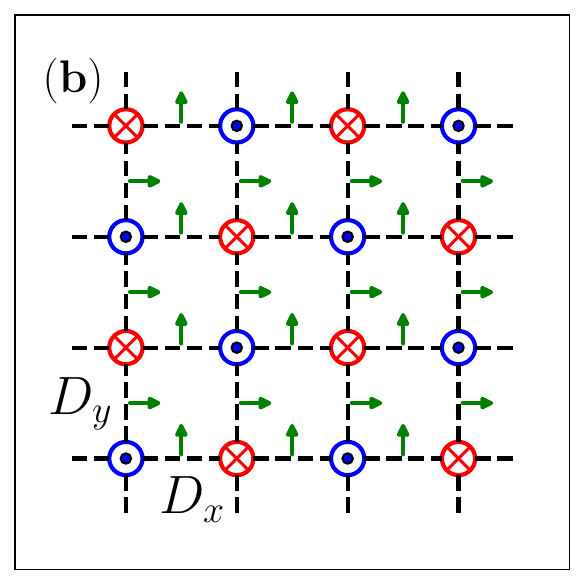}
         \captionlistentry{}
         \label{antiferromagnet}
     \end{subfigure}
     \captionsetup{skip=-2pt}
     \caption{Schematic diagram of the DM vectors in a 2D  square lattice for (a) ferromagnet and (b) antiferromagnet.}
     \label{2D lattice}
\end{figure}

The material candidates presented above are different spin-systems with DMI, and some with even anisotropic DMI, however achieving a $S=\frac{1}{2}$ Heisenberg ferro- or an antiferromagnet with an anisotropic DMI, such as ours, is still lacking. Nevertheless, motivated by the experimental and theoretical realizations of anisotropic DMI in various magnetic systems, in this work, we investigate its effects on the topological properties, and ascertain some of the key physical properties, such as, edge modes, THE etc, which are computed. There is a subtle point that requires attention. In fermionic systems, the quantized conductivity is proportional to the Chern number of the occupied bands below the Fermi energy, while in bosonic systems, there is no concept of Fermi energy. Hence, owing to no constraint in the occupancy of bosons in a particular quantum state, the lower bands are considered for computing these nontrivial transport properties at low temperatures.


We aim to study the low energy spin excitations, which are energetically separated from the continuum~\cite{magnon-pair}.~Under a spin-boson transformation, these spin excitations or the magnons, that are quanta of magnetic excitations behave like bosonic particles. In presence of an anisotropic DMI, we shall study the topological behaviour of magnons on a 2D square lattice in presence of both ferromagnetic (FM) and antiferromagnetic (AFM) correlations in order to compare and contrast between the two. 
In such ordered systems, we take the magnetization along the $z$-direction as shown in Fig.~\ref{2D lattice}. The DM vectors act perpendicular to the magnetization vectors. Although we shall deal with single-magnon and two-magnon excitations in both the FM and AFM cases, nevertheless, in the AFM case, the low-energy three- and four-magnon excitations also need to be considered that arise through second-order process (will be explained later).

Let us give a glimpse of our spin model in the presence of an external magnetic field that we study in the following. The Hamiltonian for our system comprises of a Heisenberg exchange interaction, a DMI, and a Zeeman term. It can be mapped onto a bosonic Hamiltonian, which can hence be numerically solved. To study the effect of an anisotropic DMI, we introduce a tuning parameter $\alpha=\frac{D_x}{D_y}$ and abbreviate $D_y$ = $D$. As discussed earlier, $\alpha$ can be positive or negative~\cite{S4}. We are essentially interested in studying the band structure, bulk boundary correspondence, and thermal Hall effect in presence of both (FM and AFM) kinds of magnetic ordering as a function of $\alpha$.

The paper is organised as follows. In Sec.~\ref{Formalism}, we discuss some general properties of spin-excitations, spin-boson transformation, $U(1)$ symmetry, and how the DMI breaks the space inversion symmetry (SIS) in centrosymmetric materials (such as a square-lattice). In Sec.~\ref{model} (and in Appendix~\ref{appendix:A}), we construct the model Hamiltonian for FM (Appendix~\ref{appendix_ferro}) and AFM (Appendix~\ref{appendix_antiferro}) on a 2D square lattice. In Sec.~\ref{results}, we study the behaviour of topological spin excitations in the presence of anisotropic DMI and the thermal Hall conductivity as a signature of transport phenomena for both FM and AFM cases.


\begingroup
\allowdisplaybreaks
\section{MODEL AND FORMALISM}
\label{Formalism}
\subsection{Matsubara-Matsuda transformation}
\label{Preliminaries}
\endgroup
To understand the formalism, let us define the ground state to be denoted by,
\begingroup
\allowdisplaybreaks
\begin{equation}
    \ket{0}=\bigotimes_{i=1}^N\ket{S_i^z}.
\end{equation}
\endgroup
Here, $N$ is the total number of spins, and  $S_i^z$ is the $z$ component of spin operator $S_i$.
This $\ket {0}$ state is known to be in the $\Delta S = 0$ sector. $\Delta S$ can be 0, 1, 2, 3... which labels the $z$ component of spin relative to the ground state of the system. When $S_i^z$ acts on the state $\ket{0}$, it gives the value of $z$ component of the spin angular momentum situated at the $i^{\text{th}}$ site.

For a ferromagnet, $\Delta S = 1$ sector is given by a new state,
\begin{equation}
    \ket{i}=\frac{1}{\sqrt{2S}} S_i^{-}\ket{0},\label{One_magnon_state_ferro}
\end{equation}
which is denoted as a single-magnon excitation. Now, to produce a two-magnon excitation in the $\Delta S = 2$ sector, we need to flip two spins at different positions, that can be expressed via, 
\begin{equation}
    \ket{i,j} = \frac{1}{2S}S_i^{-}S_j^{-}\ket{0}. 
\end{equation}
In case of an antiferromagnet, there are two sublattices, namely, A (occupied by $\uparrow$-spin) and B (occupied by $\downarrow$-spin). Thus, at an arbitrary $i^{\text{th}}$ location, the spin orientation could either be $\uparrow$ or $\downarrow$ depending on whether it belongs to an A or a B sublattice. When a single $\uparrow$-spin at $i^{\text{th}}$ position flips we get a single-magnon excitation, what we call a one-magnon state, which is denoted by Eq.~\eqref{One_magnon_state_ferro}. On the other hand, if a $\downarrow$-spin at $j^{\text{th}}$ position flips, the resulting one-magnon state is represented by 
\begin{equation}
    \ket{j}=\frac{1}{\sqrt{2S}} S_j^{+}\ket{0}.\label{One_magnon_state_antiferro}
\end{equation}
Further, two-magnon states can be created by flipping two spins and they can be defined by three distinct ways, namely
\begin{equation}
    \ket{i,j}=
    \begin{cases}
     \frac{1}{2S}S_i^{-}S_{j}^{-}\ket{0},~~~~\text{(a)}\\
     \frac{1}{2S}S_i^{+}S_{j}^{+}\ket{0},~~~~\text{(b)} \\
     \frac{1}{2S}S_i^{-}S_{j}^{+}\ket{0},~~~~\text{(c)} 
    \end{cases}
    \label{two-magnon1}
\end{equation}
where the $i^{\text{th}}$ and the $j^{\text{th}}$ lattice sites belong to either A or B sublattice in Eq.~(\ref{two-magnon1}(a)) or Eq.~(\ref{two-magnon1}(b) respectively, and they belong to different sublattices in Eq.~(\ref{two-magnon1}(c)). The latter is also known as a spin-zero two-magnon bound states.
In a similar way, we can also express three-magnon and four-magnon states using these spin operators, depending on whether the $i$ and $j$  belong to an A or a B sublattice. We have shown them in Appendix~\ref{appendix:three- and four-magnon}.

To solve our Hamiltonian (see Eq.~\eqref{total_H} below), in two dimensions, we shall convert the $S=\frac{1}{2}$ operator to bosonic operators ($a_i$ and $a_i^\dg$) for a ferromagnet via Matsuda-Matsubara transformation~\cite{matsubara1956}, namely,
    \begin{align}
        S_i^- &= a_i,\nonumber\\
        S_i^+ &= a_i^\dg,\nonumber\\
        S_i^z &= a_i^\dg a_i - \frac{1}{2}.
        \label{eq:Matsuda1}
    \end{align}    
While the same transformation equation is obeyed for an antiferromagnet corresponding to sites in the A sublattice, the operators in the B sublattice are converted as, 
    \begin{align}
        S_i^- &= b_i^\dg,\nonumber\\
        S_i^+ &= b_i,\nonumber\\
        S_i^z &= \frac{1}{2} - b_i^\dg b_i.  
        \label{Matsuda2}
    \end{align}

\subsection{$U(1)$ symmetry} 
\label{U1_symmetry}
We shall investigate two systems, namely, collinear ferromagnets and antiferromagnets, both of which have spins in a preferred direction. In this paper, we have considered the preferred direction as the $z$-axis, and hence consider XXZ type Heisenberg exchange interaction, where $J_z$ must be greater than both $J_x$ and $J_y$, and in addition, $J_x = J_y$, that is, the term  $H_1 = \mathbf{S}_{p,q} \cdot \mathbf{I}\,\mathbf{S}_{p+1, q}$ has the form,
\begin{equation}
    \frac{J^x}{2} \left[\left(S^+_{p,q}S^-_{p+1,q} + S^-_{p,q}S^+_{p+1,q}\right)\right] + J^z S_{p,q}^zS_{p+1,q}^z,\label{XXZ}
\end{equation}
and known as a $U(1)$ symmetric spin interaction in a ferromagnet. Hence, it will commute with the $z$-component of total spin, that is, $[H_1, S^z]=0$, which does not allow coupling between different particle sectors and gives us fully polarised exact quantum ground state. Conversely, for an antiferromagnet, the presence of $H_1$ gives rise to terms such as, $S^+_{p,q}S^-_{p+1,q}  (a^\dagger_{p,q}b^\dagger_{p+1,q}$), $S^-_{p,q}S^+_{p+1,q}(a_{p,q}b_{p+1,q})$, which are responsible for coupling two particle sectors, that differ by spin two, i.e, $\Delta S = 2$. Here, in this paper, we are concerned about the topological hybridization between the one- and the two-magnon bound states, and there is no direct coupling between $\Delta S = 1$ and $\Delta S = 2$ sectors via $H_1$. Instead, it plays a significant role through a second-order process in the effective Hamiltonian  (see Appendix \ref{appendix_antiferro}). Thus, the breaking of $U(1)$ symmetry via $H_1$ is subleading in this case.  

However, the DMI term ($H_2 \sim \mathbf{D}_{p,q} \cdot (\mathbf{S}_{p,q} \times \mathbf{S}_{p+1,q} )$) owing to a form,
\begin{equation}
\small
    D_{pq}\hat{x} \cdot \left[ \left(\frac{S_{p,q}^+ - S_{p,q}^-}{2i}\right)S_{p+1,q}^z - \left(\frac{S_{p+1,q}^+ - S_{p+1,q}^-}{2i}\right)S_{p,q}^z \right],
\end{equation}
couples the two sectors separated by one excitation and is responsible for the hybridization of one-magnon and two-magnon excitations. Hence, it breaks the $U(1)$ symmetry, thereby yielding topological edge modes.

\subsection{Breaking of SIS}
\label{DMI and symmetries}
 In general, DMI is a type of antisymmetric exchange coupling that arises as a consequence of spin-orbit coupling in a magnetic system with broken inversion
  \begin{figure}[t]
     \begin{subfigure}[!h]{0.48\columnwidth}
         \includegraphics[width=\columnwidth]{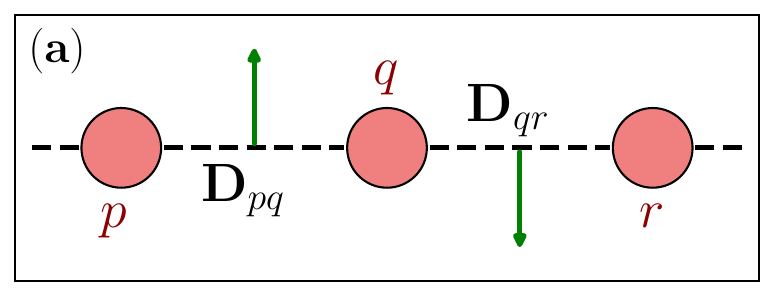}
         \captionlistentry{}
         \label{staggerd}
     \end{subfigure}
     \begin{subfigure}[!h]{0.48\columnwidth}
         \includegraphics[width=\columnwidth]{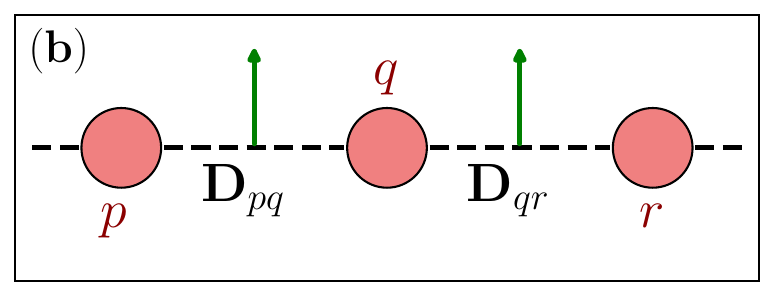}
         \captionlistentry{}
         \label{uniform}
     \end{subfigure}
     \captionsetup{skip=-2pt}
     \caption{1D model to visualize the alignment of the DM vector. (a) Staggered and (b) uniform DM interaction, where the DM vectors are perpendicular to the spin angular momentum situated at site $p$, $q$, and $r$.}
     \label{model1}
\end{figure}
 symmetry.~However, in our case, we have considered a two-dimensional square lattice 
 model with the inversion symmetry unharmed. Therefore, we must define the direction of the DM vectors in a way that implies the existence of broken inversion symmetry in the system. Here, in Fig.~\ref{model1}(a), they are situated along the bond in a staggered manner, whereas, in Fig.~\ref{model1}(b), all the DM vectors point in the same direction. In these configurations, the DMI Hamiltonian can be written as,
\begin{equation}
    H_{\text{\tiny{DMI}}} = \mathbf{D}_{pq}\cdot (\mathbf{S}_p \times \mathbf{S}_q) + \mathbf{D}_{qr}\cdot (\mathbf{S}_q \times \mathbf{S}_r).
\end{equation}
Here, $p$ and $r$ denote sites that are symmetrically placed with respect to $q$. If the inversion center is situated at the lattice site $q$, following inversion, the Hamiltonian should be given by,
\begin{equation}
    H_{\text{\tiny{DMI}}}' = \mathbf{D}_{pq}\cdot (\mathbf{S}_r \times \mathbf{S}_q) + \mathbf{D}_{qr}\cdot (\mathbf{S}_q \times \mathbf{S}_p).
\end{equation}
Presence of SIS demands that $H_{\text{\tiny{DMI}}} = H_{\text{\tiny{DMI}}}'$~\cite{kawano2019thermal}, which is only possible if $\mathbf{D}_{pq}=-\mathbf{D}_{qr}$, a scenario that is satisfied in Fig.~\ref{model1}(a), while Fig.~\ref{model1}(b) suggests breaking of SIS. Hence, we can choose all the $\mathbf{D}_x$ and $\mathbf{D}_y$ vectors to be pointing uniformly (not staggered) acted along the $x$- and $y$-directions respectively of the 2D square lattice (see Fig.~\ref{2D lattice}).
That is, we have considered the DM vector configurations to be as shown in Fig.~\ref{model1}(b).
\subsection{Hamiltonian}
\label{model}
In the following, we write down the Hamiltonian to analyze the topological excitations in presence of both FM and AFM interactions. It can be fragmented as,
\begin{equation}
    H=H_{\text{\tiny{DMI}}} + H_{\text{\tiny{H}}} + H_{\text{\tiny{Z}}}.
    \label{total_H}
\end{equation}
The first term is due to the effect of the DMI term, and for our system, it is given by,
\begin{equation}
    H_{\text{\tiny{DMI}}}=\sum_{p=1}^{N_x}\sum_{q=1}^{N_y} [ D \hat{x} \cdot (\mb{S}_{p,q} \times \mb{S}_{p,q+1}) + \alpha D \hat{y} \cdot (\mb{S}_{p,q} \times \mb{S}_{p+1,q})],\label{DMI_TRS}
\end{equation}
where it may be noted that the DMI in $x$- and $y$-directions are taken as $\alpha D$ and $D$, respectively, where $\alpha$ is a tuning parameter that quantifies the anisotropy present in the system. However, the second and the third terms include the Heisenberg exchange interaction and the Zeeman terms respectively, which can be written as
\begin{align}
    H_{\text{\tiny{H}}} = \sum_{p=1}^{N_x}\sum_{q=1}^{N_y} &[\mb{S}_{p,q} \cdot \mb{I}_1 (\mb{S}_{p+1,q} + \mb{S}_{p,q+1})\nonumber\\
    &+ \mb{S}_{p,q}\cdot \mb{I}_2 (\mb{S}_{{p+2},q} + \mb{S}_{p,q+2}) ],\nonumber\\
    H_{\text{\tiny{Z}}}=\sum_{p=1}^{N_x}\sum_{q=1}^{N_y}& -B_0S_{p,q}^z.
\end{align}
Here, $\mb{I}_1$ and $\mb{I}_2$ are XXZ type exchange interactions~\cite{XXZ2022} with $\mb{I}_1 = \text{diag}(J_1,J_1, J_1^z)$ for the nearest neighbours (NN) and $\mb{I}_2=\text{diag}(J_2,J_2,J_2^z)$ for next to next nearest neighbours (NNNN). To realise two-magnon bound states separated from the two-magnon continuum, we need to consider a longer range, that is, the NNNN term.~We shall study the topological response of low-magnon excitations in our system in the presence of an anisotropic DMI.

To observe topological gaps, breaking of time reversal symmetry (TRS) is essential. In a magnetic system, the time reversal operator $\tau$ flips the direction of the spin ($\uparrow \rightarrow \downarrow$ and vice versa). Hence, any magnetically ordered ground state violates this symmetry. Whereas, effective TRS is a good symmetry for the magnetic system and acts differently for FM and AFM cases.

In case of a ferromagnetic square lattice, effective TRS is defined by the combination of the time reversal operator $\tau$ and a rotational operator $R(n, \pi)$, where $n$ being the direction perpendicular to that of the ordered state magnetization. Here, the spins are polarised along the $z$-direction, hence, $n$ can be $x$ or $y$. The operator $R(n, \pi)$ performs a rotation by $\pi$ in spin space about the $n$-axis. Consequently, despite flipping the spin via the $\tau$ operator, it preserves its ground state orientation obtained through the $R(n, \pi)$ operator, thereby maintaining spin-space isotropy. However, this rotation affects the DM vectors ($D_x$, $D_y$), causing them to change sign, while they do not change sign under the time reversal operator. Thus, it breaks the effective TRS as well as spin-space isotropy to gap out the energy bands. It needs to be mentioned that the external magnetic field $B_0$ is not capable of breaking the effective TRS and is only responsible for shifting the magnon energies, as can be seen by analyzing the effective Hamiltonian in reciprocal space. As a result, we shall get a non-zero anomalous thermal Hall conductivity~\cite{anomalousTHE1,anomalousTHE2,anomalousTHE3} even in the absence of $B_0$ as seen in subsequent discussions. However, for simplicity, we shall continue to refer to it as the thermal Hall conductivity throughout our discussion.

Distinct from the FM case, in an antiferromagnetic square lattice, the effective TRS is defined as a $t\tau$ symmetry, where the the time reversal operator $\tau$ flips the spin orientation and $t$ is the translation connecting a site of one sublattice to another. In absence of $B_0$, for $D \neq 0$, the system still obeys effective TRS. Only the presence of $B_0$ can destroy the effective TRS, as the sublattice translation ($t$) will be violated due to the difference of local Zeeman energies. Hence, the thermal Hall conductivity should be zero in the absence of an external magnetic field. In the following discussion, we shall explain how both the DMI and $B_0$ are responsible for observing a non-zero thermal Hall conductivity in the AFM case.

To relate the band dispersion and the topological properties of single-magnon and two-magnon bound states, we need to establish an effective three-band model from the Hamiltonian given in Eq.~\eqref{total_H}. We can write the Hamiltonian in the basis of the one-magnon and two-magnon states whose corresponding details are presented in Appendix \ref{appendix:A}. The Hamiltonian for a ferromagnet in the momentum space, namely $H_{\text{\tiny{F}}}(\mathbf{k})$ can be written as (see Eqs.~\ref{hopping_ferro},~\ref{R}), 

\begin{equation}
    H_{\text{\tiny{F}}}(\mathbf{k})=\begin{pmatrix}
                    E_{s_{\tiny{\text{F}}}} & -i\alpha D \sin{\frac{k_x}{2}} & D\sin{\frac{k_y}{2}}\\
                    i\alpha D \sin{\frac{k_x}{2}} & E_{t_{\text{\tiny{F}}}}^{x,y} & P\\
                    D\sin{\frac{k_y}{2}} & P & E_{t_{\text{\tiny{F}}}}^{y,x}
                \end{pmatrix},\label{effective_ferro}
\end{equation}
where, $P = \frac{2J_1^2}{J_1^z}\cos(k_x/2)\cos(k_y/2)$, $E_{s_{\tiny{\text{F}}}}$ stands for single-magnon dispersion, $E_{t_{\text{\tiny{F}}}}^{x,y}$ and $E_{t_{\text{\tiny{F}}}}^{y,x}$ are related to the two-magnon energies, and they are given by,
\begin{subequations}
 \begingroup
 \allowdisplaybreaks
\begin{align}
     E_{s_{\tiny{\text{F}}}} = B_0 &- 2 (J_1^z + J_2^z) + J_1 (\cos{k_x} + \cos{k_y})\nonumber\\
     &+ J_2[\cos{(2k_x)} +\cos{(2k_y)} ],\\
     E_{t_{\text{\tiny{F}}}}^{p,q} = 2B_0 &-3J_1^z - 4J_2^z + J_2 \cos{k_p} + \frac{J_1^2}{J_1^z}\cos^2{\frac{k_p}{2}}\nonumber\\
     +\frac{J_2^2}{J_1^z}&\cos^2{k_p} + 2\frac{J_1^2}{J_1^z}\cos^2{\frac{k_q}{2}} + 2\frac{J_2^2}{J_1^z}\cos^2{k_q}.
\end{align}   
\endgroup
\end{subequations}
Here $(p,q)=(x,y)$ and it should be noted that $J_1,~J_1^z$ are negative for the ferromagnetic interaction among neighbouring spins, but the NNNN interaction will be antiferromagnetic so that the two-magnon bound states become energetically favourable. In the absence of DMI, Eq.~\eqref{effective_ferro} does not break the $U(1)$ symmetry as well as the effective TRS, namely, $H(-\mathbf{k}) = H^*(\mathbf{k})$. However, in the presence of a DMI, it breaks the spin conservation as well as the effective TRS, which, as we shall see, may induce a topological phase transition.

 In a similar manner, we obtain the effective Hamiltonian for an antiferromagnet ($H_{\text{\tiny{AF}}}(\mathbf{k})$) from Eqs.~(\ref{hopping_anti},~\ref{R2}), which can be written as,
\begin{equation}
    \footnotesize
    H_{\text{\tiny{AF}}}(\mathbf{k})=
                    \begin{pmatrix}
                        E_{s_\text{\tiny{AF}}}^1 & 0 & -f(k) & f(-k) & - g(k) & g(-k)\\
                        0 & E_{s_\text{\tiny{AF}}}^2 & -f(-k) & f(k) & g(-k) & - g(k)\\
                        -f^*(k) & -f^*(-k) & E_{t_\text{\tiny{AF}}}^x & u_2(k) & u_4(k) & u_5(k)\\
                        f^*(-k) & f^*(k) & u_2(k) & E_{t_\text{\tiny{AF}}}^x & u_5(k) & u_4(k)\\
                        - g^*(k) & g^*(-k) & u_4(k) & u_5(k) & E_{t_\text{\tiny{AF}}}^y & u_3(k)\\
                        g^*(-k) & - g^*(k) & u_5(k) & u_4(k) & u_3(k) & E_{t_\text{\tiny{AF}}}^y
                    \end{pmatrix},\label{effective_antiferro}    
\end{equation}
where $E_{s_\text{\tiny{AF}}}^1$ and $E_{s_\text{\tiny{AF}}}^2$ represent one-magnon excitations, while $E_{t_{\text{\tiny{AF}}}}^n$ with $n=x,y$ are related to the two-magnon energies along $x$- and $y$-directions, and they are given by,
\begin{subequations}
    \begin{align}
        E_{s_\text{\tiny{AF}}}^1 =  2J_1^z &- 2 J_2^z + B_0 - \frac{J_1^2}{J_1^z + B_0} (\cos{k_x} + \cos{k_y})\nonumber\\ 
        &+ 2\left(J_2 - \frac{J_1^2}{2J_1^z + 2B_0}\right) \cos{k_x}\cos{k_y} ,\\
        E_{s_\text{\tiny{AF}}}^2 =  2J_1^z &- 2 J_2^z - B_0 - \frac{J_1^2}{J_1^z - B_0} (\cos{k_x} + \cos{k_y})\nonumber\\ 
        &+ 2\left(J_2 - \frac{J_1^2}{2J_1^z - 2B_0}\right) \cos{k_x}\cos{k_y} ,\\
        E_{t_{\text{\tiny{AF}}}}^x =  3J_1^z &- 4J_2^z -\frac{J_1^2}{2 J_1^z} - \frac{3J_2^2}{2 J_1^z} - \frac{J_2^2}{2J_1^z} \cos{(k_x - k_y)}\nonumber\\
        &- \frac{J_2^2}{J_1^z} \cos{(k_x + k_y)}\\
         E_{t_{\text{\tiny{AF}}}}^y =  3J_1^z &- 4J_2^z -\frac{3J_1^2}{2 J_1^z} - \frac{3J_2^2}{2 J_1^z} - \frac{J_2^2}{2J_1^z} \cos{(k_x + k_y)}\nonumber\\
        &- \frac{J_2^2}{J_1^z} \cos{(k_x - k_y)}
    \end{align}
\end{subequations}
Other matrix elements in Eq.~\eqref{effective_antiferro} are given by, 
\begin{subequations}
 \begingroup
 \allowdisplaybreaks
\begin{align}
    f(k)&=\frac{D_x}{2} e ^ {i(\frac{k_x}{4} - \frac{k_y}{4})},~~g(k) = - i\frac{D_y}{2} e ^ {i(\frac{k_x}{4} + \frac{k_y}{4})}\\
    u_2(k) &=J_2\cos{\left(\frac{k_x}{2} - \frac{k_y}{2}\right)} - \frac{J_1^2}{2J_1^z}\cos{\left(\frac{k_x}{2} + \frac{k_y}{2}\right)}\\
    u_3(k) &=J_2\cos{\left(\frac{k_x}{2} + \frac{k_y}{2}\right)} - \frac{J_1^2}{2J_1^z}\cos{\left(\frac{k_x}{2} - \frac{k_y}{2}\right)}\\
    u_4(k) &= - \frac{J_1^2}{2J_1^z}\cos{\frac{k_y}{2}},~~ u_5(k) = - \frac{J_1^2}{2J_1^z}\cos{\frac{k_x}{2}}.
    \end{align}
  \endgroup  
\end{subequations}

Similar to the ferromagnetic case, we assume that all the exchange couplings, namely, $J_1^z$, $J_1$, $J_2^z$ and $J_2$ to be positive, so that two-magnon bound states become energetically favourable. By breaking the effective TRS and the spin conservation of the system, we shall study topological phenomena in the reciprocal space. In the following section, we shall present the results on the band structure and the topological properties for both FM and AFM couplings and at the end, make a comparison between the two.


\section{RESULTS AND DISCUSSION}
\label{results}
\subsection{Magnons in a Ferromagnet}
\label{results_ferro}
It is important to realise and stated earlier that for a ferromagnetic Hamiltonian on a square lattice, the energy penalty due to single-magnon (one-magnon) excitation is smaller than the two-magnon bound states. Further, both the effective TRS and the SIS are violated in presence of a DMI. 
\begin{figure}[b]
\centering
    \includegraphics[width=1\columnwidth]{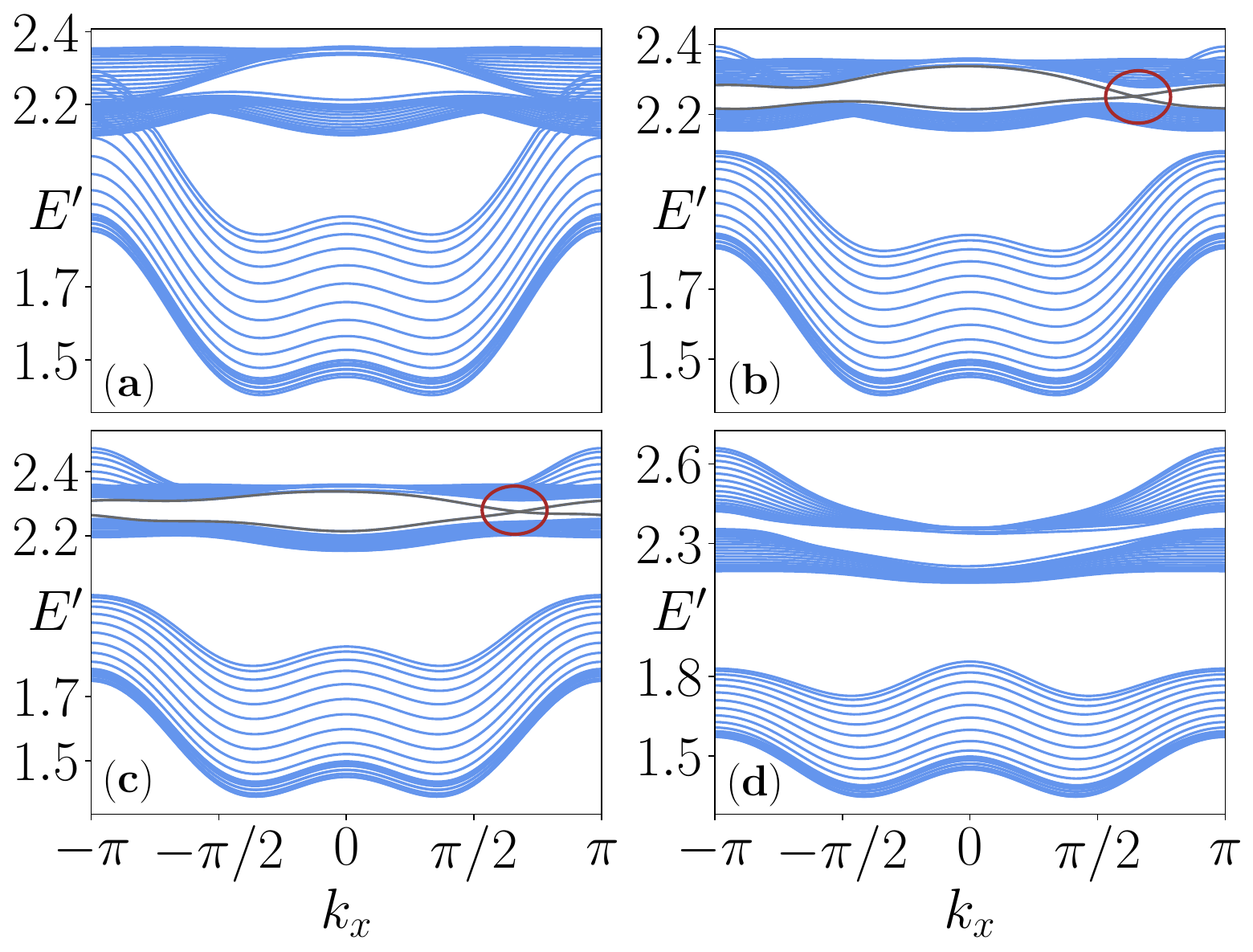}
    \caption{Topological spin excitations due to one-magnon and two-magnon bound states in the presence of an anisotropic DMI give rise to the chiral edge state. (a) Without DMI, we get bands touching at the end of the BZ, which is a trivial state. For $D/J^z_1=-0.1$, if we take (b) $\alpha=1$ or (c) $\alpha = 2$, the presence of a chiral edge state between the gaped two-magnon bound state conveys the topological state. Whereas (d) $\alpha = 4$ gives no chiral edge state to signify the topological state. Values for other parameters follow $J_1/J_1^z = 0.2$, $J_2/J_1^z = -0.1$, $J_2^z/J_1^z=-0.3$, $B/|J_1^z| = 0.3$. Band energy is scaled here, with $E'=E/|J_1^z|$.}
    \label{ribbon}
\end{figure}
In Fig.~5 of Ref.~\cite{magnonboundpairs2023}, in the absence of DMI, the energy bands for the one-magnon and two-
magnon bound states touch each other at the corners of the Brillouin zone (BZ), and is indicative of the existence of a trivial phase under effective TRS and $U(1)$ symmetry. Whereas, violation of these symmetries results in the opening of a band gap.
It needs to be ascertained whether the gap is topological or trivial. We take recourse to the bulk boundary correspondence considering a semi-infinite ribbon with a finite number of lattice sites along the $y$-direction (while the ribbon is infinite along the $x$-direction). The presence of two edge states that are chiral partners, and exist in the spectral gap testifies the topological nature of the system. In the presence of an anisotropic DMI, with $\alpha$ as an anisotropy parameter, we plot the dispersion spectrum for a semi-infinite ribbon, as shown below in Fig.~\ref{ribbon}.
Here, we have considered two distinct values of $\alpha$, namely, $\alpha=2$ and $\alpha=4$. These are just representative values, and in principle, any value of $\alpha$ ($\neq1$) may be used. However, smaller values of $\alpha$ do not give rise to any observable consequences, at least on the topic we are interested in. We observe presence of 
the chiral edge modes for $\alpha = 2$, while these modes are absent for larger $\alpha$, namely, $\alpha = 4$. In
particular, for an intermediate value, that is, $\alpha =3.6$ \cite{note}, we observe closing of the spectral gap. This implies that there is a topological phase transition occurring at $\alpha=3.6$. The touching of the bands for the two magnon sectors is depicted in Fig.~\ref{gapclossing}(a).~The other value of $\alpha$, namely $\alpha=4$ does not demonstrate a topological phase and is characterized by no chiral edge modes, as discussed earlier.

\begin{figure}[t]
\centering
    \includegraphics[width=1\columnwidth]{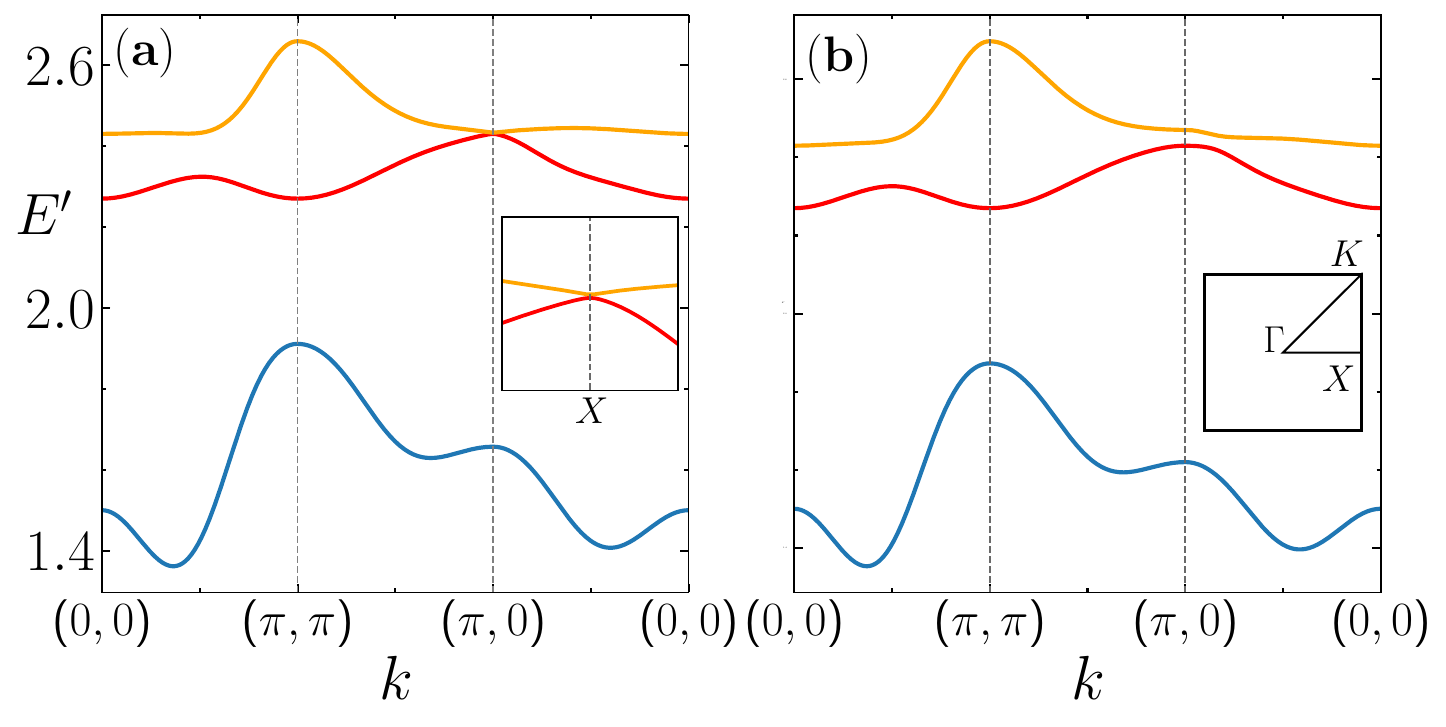}
    \caption{Band dispersion ($E'$) for one-magnon and two-magnon bound states for the ferromagnetic square lattice shows the gap-closing transition in the presence of anisotropic DMI. (a) In the presence of anisotropic DMI with $D/J_1^z=-0.1$ and $\alpha=3.6$, Two bands touch at $X$ point inside the BZ, (b) but for $\alpha=4$ the bands remain gapped.}
    \label{gapclossing}
\end{figure}
In addition to the presence of chiral edge states, the topological phase of this effective model can be identified by the
non-zero values of the Berry curvature corresponding to these three (one one-magnon and two two-magnon) bands. The Berry curvature of the $n^{\text{th}}$ band is given by,
\begin{equation}
    \Omega_{xy}^n = -2\text{Im}\braket{\partial_{k_x}\mathbf{u}_n(\mathbf{k})|\partial_{k_y}\mathbf{u}_n(\mathbf{k})}.
\end{equation}
Here, $n$ is the band index, $\ket{\mathbf{u}_n(\mathbf{k})}$ is the $n^{\text{th}}$ eigenvector corresponding to the  $n^{\text{th}}$ energy band and $\mathbf{k}$ is the wavevector.
The topological invariant, namely, the Chern number ($C_n$) is obtained via integrating the Berry curvature over the BZ~\cite{thouless1998topological}, where $n$ is the band index,
\begin{equation}
    C_n = \frac{1}{2\pi}\int_{BZ}\mathbf{\Omega}_n(\mathbf{k})\cdot d\mathbf{k}.\label{Chern}
\end{equation}

By computing the Berry curvature corresponding to those three bands on either side of the transition point ($\alpha\simeq3.6$), it becomes evident that these bands acquire non-zero Berry curvature, as the effective TRS still remains broken beyond the transition. However, irrespective of having non-zero Berry curvature, the lower band always corresponds to a zero Chern number.

Having calculated the Chern numbers corresponding to these three bands, denoted by 0, -1, 1 in going from bottom to top of the spectrum respectively, we can substantiate our findings in Fig.~\ref{ribbon}, where two chiral edge states are observed 
between the middle and the upper bands, reflecting the difference in the Chern numbers of the adjacent bands. 
Further, it 
is important to note that we observe the absence of nontrivial edge states between the lower and middle bands as well. The lower band does not undergo band inversion during the phase transition. It may be pertinent to mention that in a kagome
lattice (Ref.~\cite{mook2014magnon}), one observes three bands with Chern numbers -1, 0, and 1. All the three bands undergo
band inversion during the course of the phase transition, where edge states are found to traverse between any two bands~\cite{mook2014edge}.
\begin{figure}[t]
         \centering
         \includegraphics[width=0.85\columnwidth]{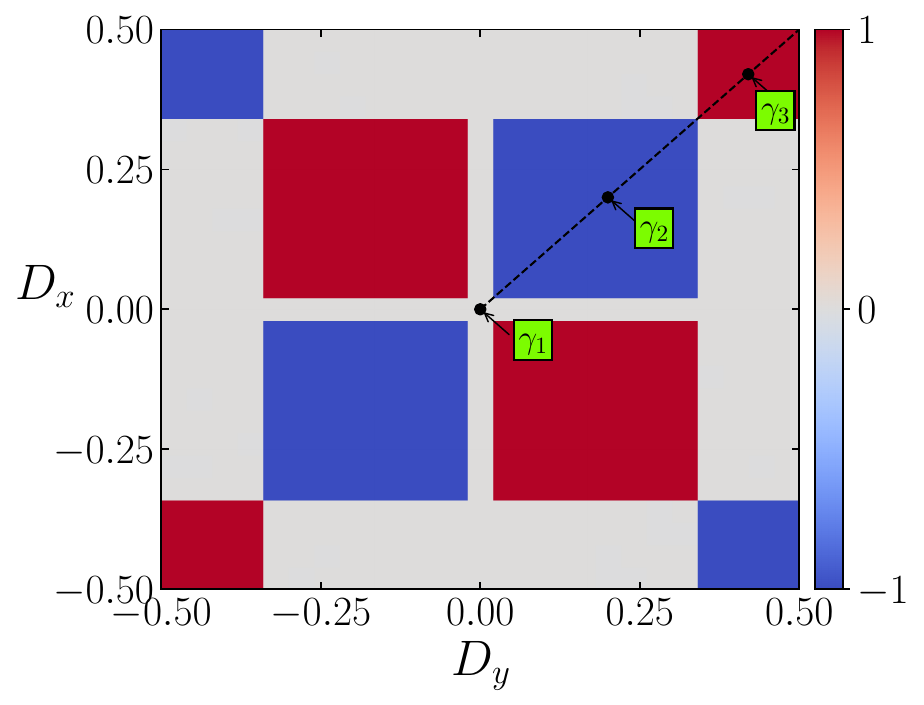}
         \caption{Chern number phase diagram of the lower band of two-magnon bound states in the ferromagnetic square lattice. The black dotted line denotes $\alpha=1$ line. There are three points $\gamma_1, \gamma_2$ and $\gamma_3$ situated at three different phases.}
         \label{phase_plot}
   \end{figure}

To further visualize the topological and trivial phases arising in our system, we construct a phase diagram in a parameter space spanned by $D_x$(= $\alpha D$) and $D_y$(= $D$) based on the value of the Chern number corresponding to the lower band of the two-magnon spectrum (the middle of the three bands), as shown in Fig.~\ref{phase_plot}~\cite{note1}.
Here, we vary $D_x$ and $D_y$ (in unit of $J_1^z$) in a reasonable range, namely, [-0.5, 0.5].
From the phase diagram, it can be seen that the grey
region denotes a trivial phase. Further, there are multiple topological phase transitions as can be seen clearly. For example, there are transitions from topological to topological ($C=\pm1\rightarrow C= \mp1$) phases corresponding to ($|D_x|$, $|D_y|$) = (0.36, 0.36) along $\alpha=1$ line and topological to trivial phases ($C=\pm 1 \rightarrow C=0$) etc.

The corresponding band structure plots are shown at three distinct points marked as $\gamma_1$, $\gamma_2$, $\gamma_3$  along $\alpha = 1$ line. We have shown band structures for these three marked points, where for $\gamma_1$ in the region of trivial phase, the bands touch each other at the corner of the BZ, whereas, for $\gamma_2$ and $\gamma_3$, the bands are gapped.
However, both the transitions from one topological phase to another distinct topological phase ($C=\pm1 \rightarrow C=\mp1$), and a topological phase to a trivial phase ($C=\pm1 \rightarrow C=0$) are accompanied by a gap-closing scenario.

These one-magnon and two-magnon bound states behave like quasiparticles and are responsible for the trans-
port properties that are linked with their topological 
features. In a fermionic system, the electrons occupy the energy states up to the Fermi energy, giving rise to the quantized Hall conductivity. Bosonic excitations or magnons can also exhibit similar topological transport phenomena in the presence of DMI. These charge-neutral excitations produce thermal Hall conductivity at finite temperatures. The
contribution from the lower band of the two-magnon bound state is predominant in this regard, as illustrated earlier. The expression for the thermal Hall conductivity can be written as~\cite{thermalhall},
\begin{widetext}
\begin{minipage}{\linewidth}
\begin{figure}[H]
    \centering
    \includegraphics[width=0.83\columnwidth]{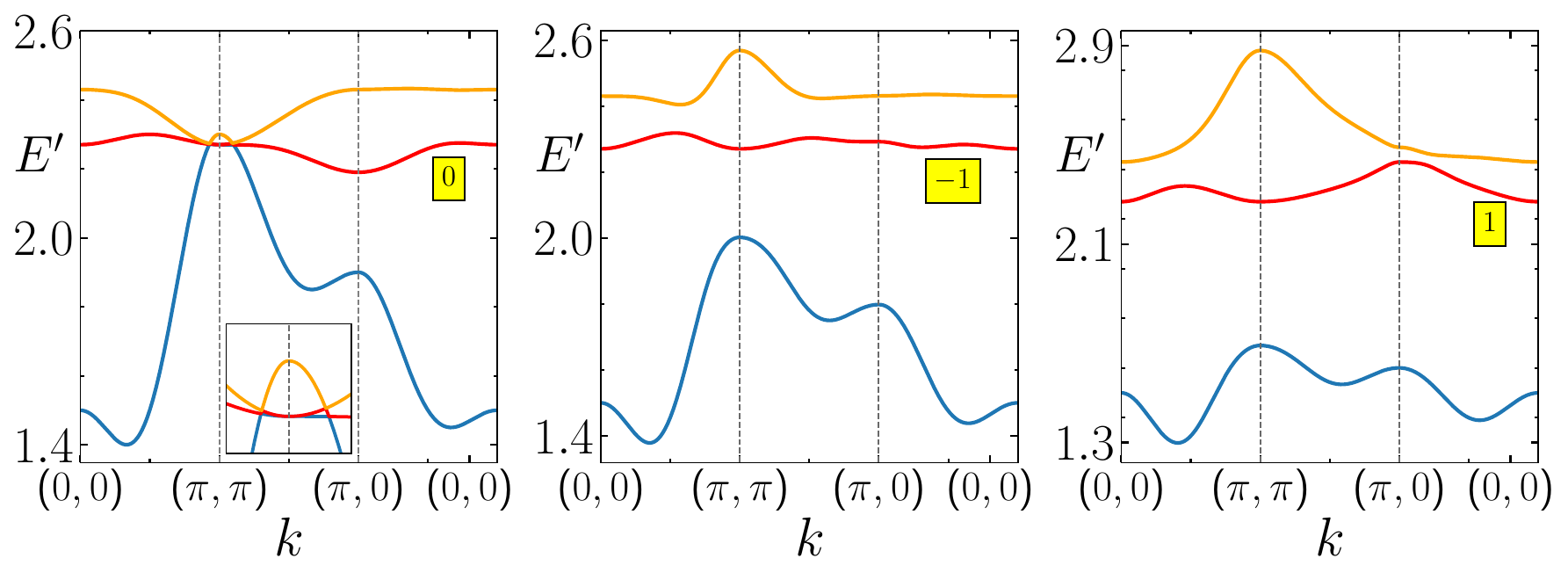}
    \caption{The band structures for $\gamma_1,~\gamma_2~\text{and}~\gamma_3$ points at three different phases along $\alpha = 1$ line have been shown. Band structure for (a) $\gamma_1$ ($D = 0$), (b) $\gamma_2$ ($D = 0.2$), (c) $\gamma_3$ ($D = 0.42$).}
    \label{fig:enter-label}
\end{figure}
\end{minipage}
\end{widetext}
\begin{equation}
    \kappa_{xy} = \frac{k_B T}{4 \pi^2} \sum_n \int_{BZ} c_2(\rho_{n,\mathbf{k}})\Omega_{xy}^n(\mathbf{k})d^2k, \label{kappa}
\end{equation}
where, $T$ is absolute temperature, $\rho_{n,\mathbf{k}} = 1 / [\exp(\varepsilon_n(\mathbf{k})/k_BT)-1]$ is the Bose-Einstein (BE)
 distribution function, $\varepsilon_n(\mathbf{k})$ being the energy of $n^\text{th}$ band, with 
\begin{equation}
\begin{split}
    c_2(\rho_{n,\mathbf{k}})=(1 + \rho_{n,\mathbf{k}})&\left(\log{\frac{1 + \rho_{n,\mathbf{k}}}{\rho_{n,\mathbf{k}}}}\right)^2  \\
    &- (\log{x})^2 - 2\text{Li}_2 (- \rho_{n,\mathbf{k}}).   
    \end{split}
\end{equation}
Here, $\text{Li}_2(-\rho_{n,\mathbf{k}})$ is a polylogarithmic function~\cite{Li2} and the thermal Hall conductivity is scaled with $\frac{\hbar}{k_B}$ and cal-
culated as a function of $\alpha$ using Eq.~\eqref{kappa} for a fixed value of $D$, shown in Fig.~\ref{Thermal hall}. 
We choose a particular value of 
$D$, namely, $D/J_1^z =-0.1$ and two distinct values for temperature ($k_BT/|J_1^z|=0.4,~0.5$). We observe the sign of the Hall conductivity to trace out the topological and trivial phase corresponding to the lower band of the two-magnon bound states. Furthermore, its magnitude increases with increasing temperature. Also, for $|\alpha|>3.6$ we get nearly zero thermal Hall conductivity.
\begin{figure}[t]
         \centering
         \includegraphics[width=1\columnwidth]{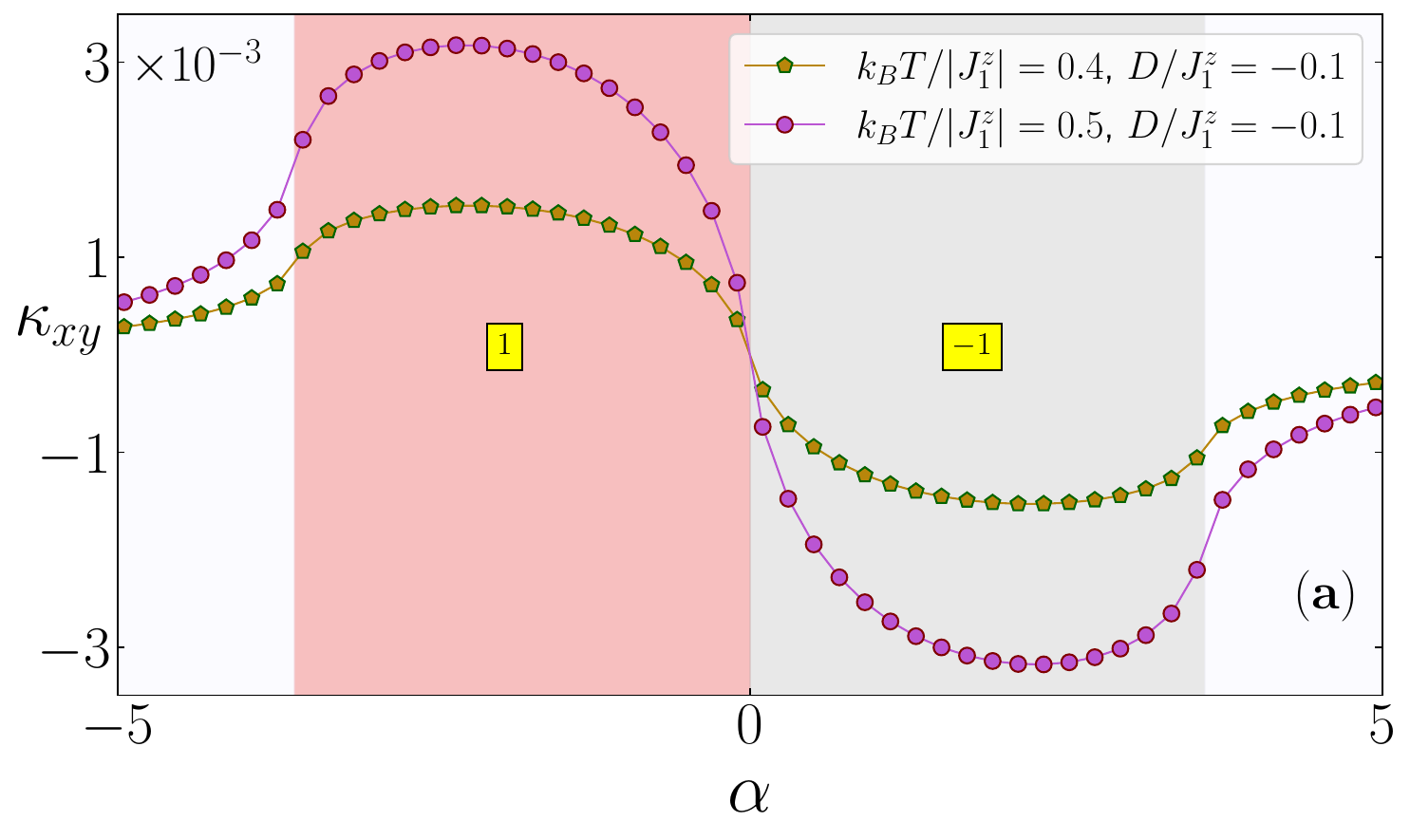}
     \caption{(a)The behaviour of the thermal Hall conductivity as a function of $\alpha$. The labels 1 and -1 denote the region with Chern number  $\pm 1$ of the lower band of two-magnon bound states.}
     \label{Thermal hall}
\end{figure}

So far we have examined the scenario for a fixed value of $B_0$, namely, $B_0=0.3$ (in unit of $J_1^z$). However, magnetic field is usually a tunable quantity in experiments. Thus, it is imperative to discuss the transition between phases as a function of the magnetic field and the role of the DMI therein. Theoretically, it is challenging to find a direct connection between the magnon band topology and an external magnetic field, since the latter does not break any relevant symmetry of the system.

\begin{figure}[t]
         \centering
         \includegraphics[width=1\columnwidth]{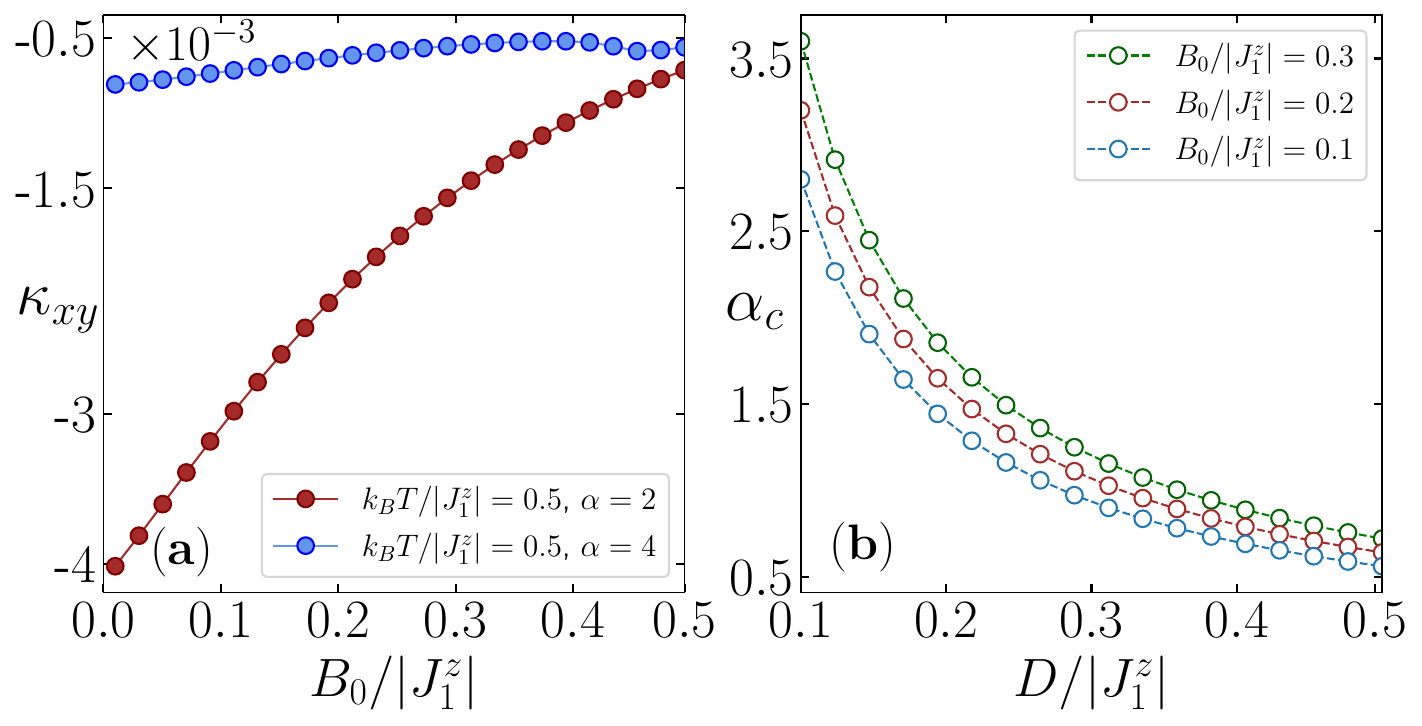}
     \caption{In the ferromagnetic case, (a) behaviour of thermal Hall conductivity as a function of $B_0$ and (b) dependency of $\alpha_c$ with $D$, for different values of $B_)$. }
     \label{kappa_vs_B_ferro}
\end{figure}

Nevertheless, to reflect upon the roles played by the external field and the DMI in study of topological phase transitions, we turn to calculating the transport properties. The thermal Hall conductivity $\kappa_{xy}$ is plotted in Fig.~\ref{kappa_vs_B_ferro}(a) as a function of $B_0$ for two different values of $\alpha$, namely, $\alpha > \alpha_c$ ($\alpha_c$: critical anisotropy) and $\alpha < \alpha_c$. This value of $\alpha_c$ ($=3.6$), of course, corresponds to $B_0=0.3|J_1^z|$. $\kappa_{xy}$ shows a sharper variation for lower $\alpha$ ($\alpha=2$), while it is fairly insensitive for larger $\alpha$ ($\alpha=4$). Empirically, we find that $\alpha_c$ increases with increasing $B_0$ (antiferromagnet shows a reverse trend as discussed later). A physical explanation for the above result can be given as follows. A larger $B_0$ aids in maintaining an ordered ground state, and is thus resilient to the formation of spin spiral states.
Further, $\alpha_c$ as a function of DMI, plotted in Fig.~\ref{kappa_vs_B_ferro}(b) reveals $\alpha_c \propto \frac{1}{D}$ for the three values of the external field, namely, $B_0/|J_1^z|=0.1,~0.2,~0.3$ shown here. Also, $\alpha_c$ is noted to be larger for higher values of $B_0$, as stated earlier.


\subsection{Magnons in an Antiferromagnet}
\label{results_antiferro}
\begin{figure}[t]
\centering
     \begin{subfigure}[!h]{0.49\columnwidth}
         \includegraphics[width=\columnwidth]{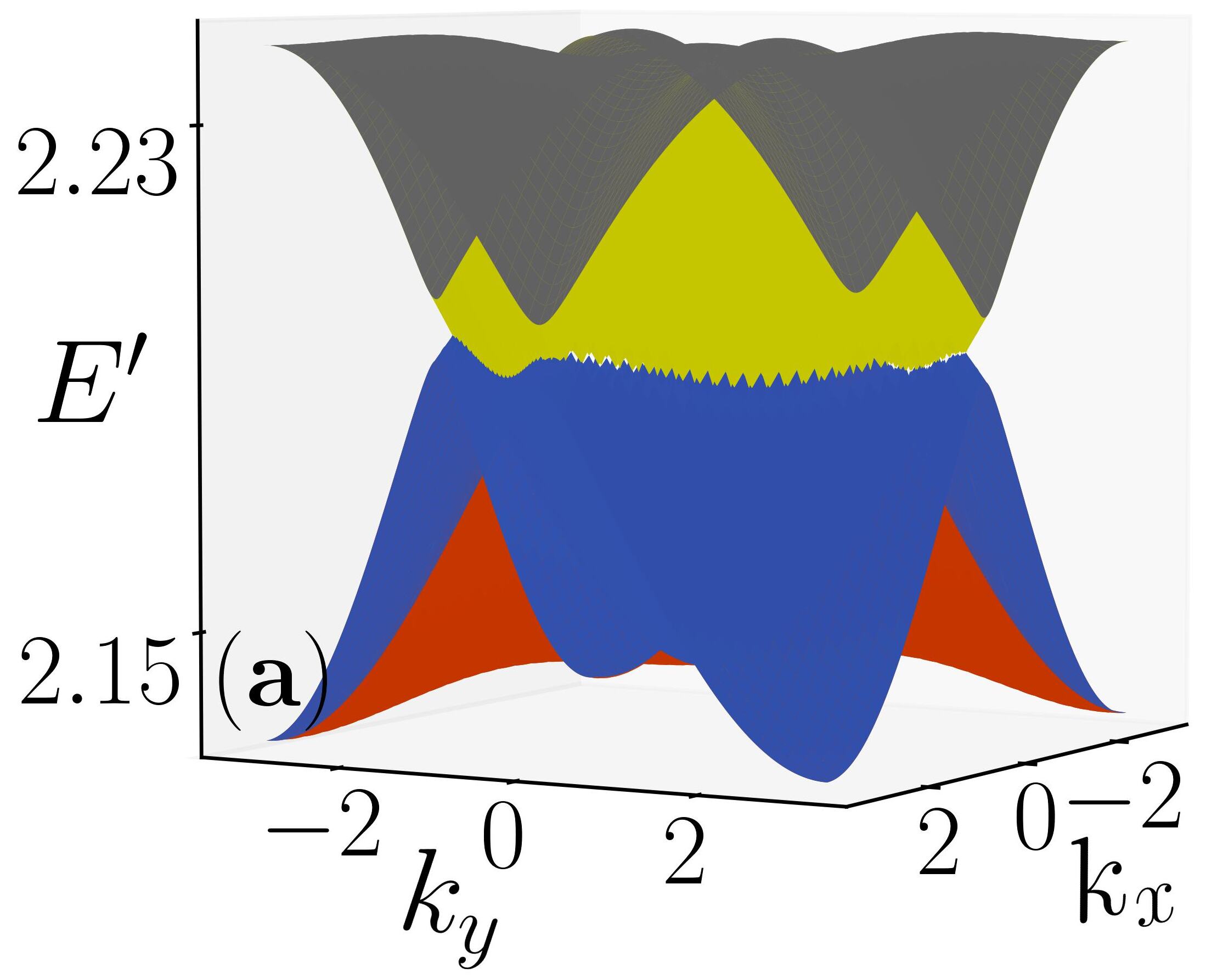}
         \captionlistentry{}
     \end{subfigure}
     \hspace{-0.2cm}
     \begin{subfigure}[!h]{0.49\columnwidth}
         \includegraphics[width=\columnwidth]{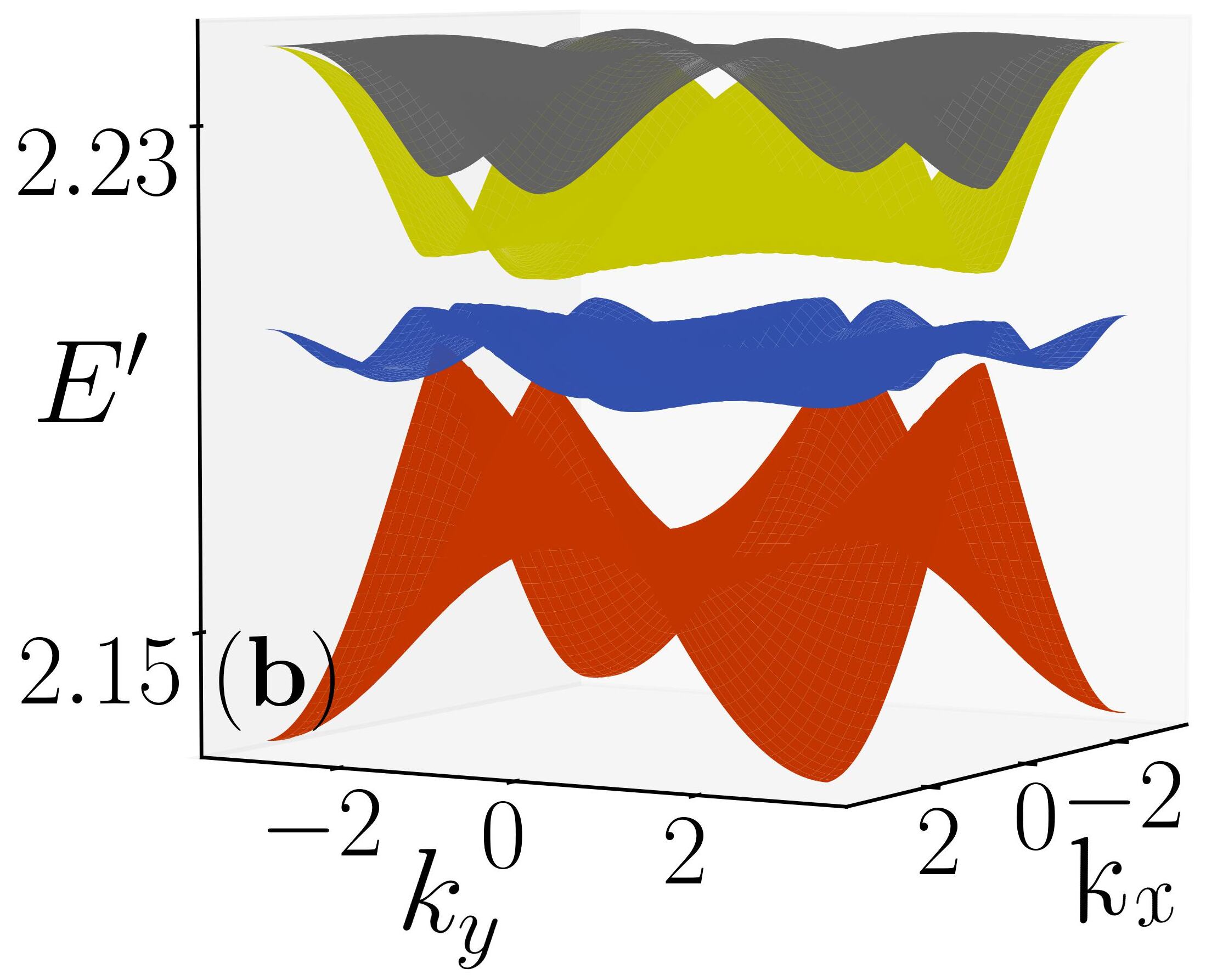}
         \captionlistentry{}
     \end{subfigure}
     \begin{subfigure}[!h]{0.49\columnwidth}
         \includegraphics[width=\columnwidth]{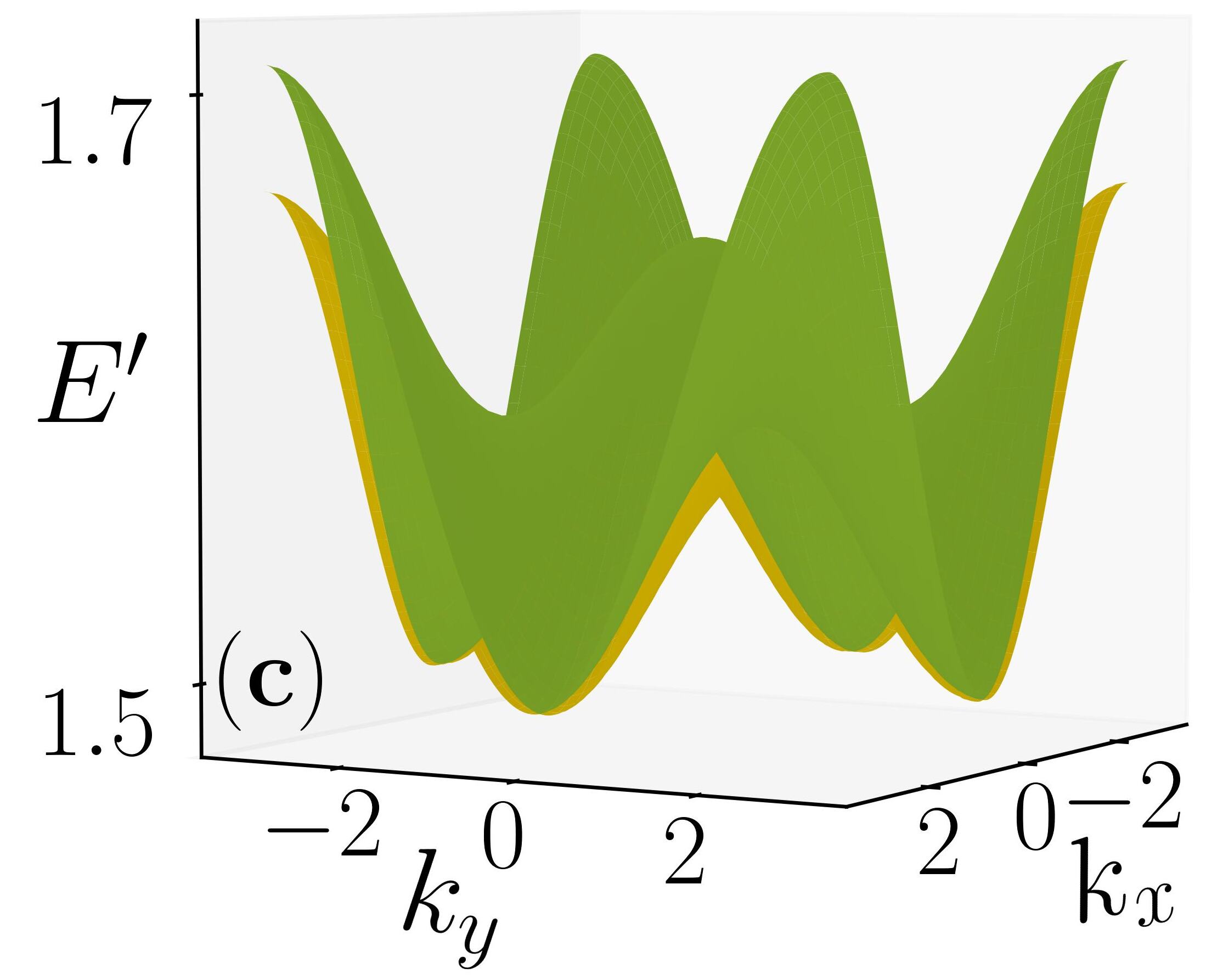}
         \captionlistentry{}
     \end{subfigure}
     \begin{subfigure}[!h]{0.49\columnwidth}
         \includegraphics[width=\columnwidth]{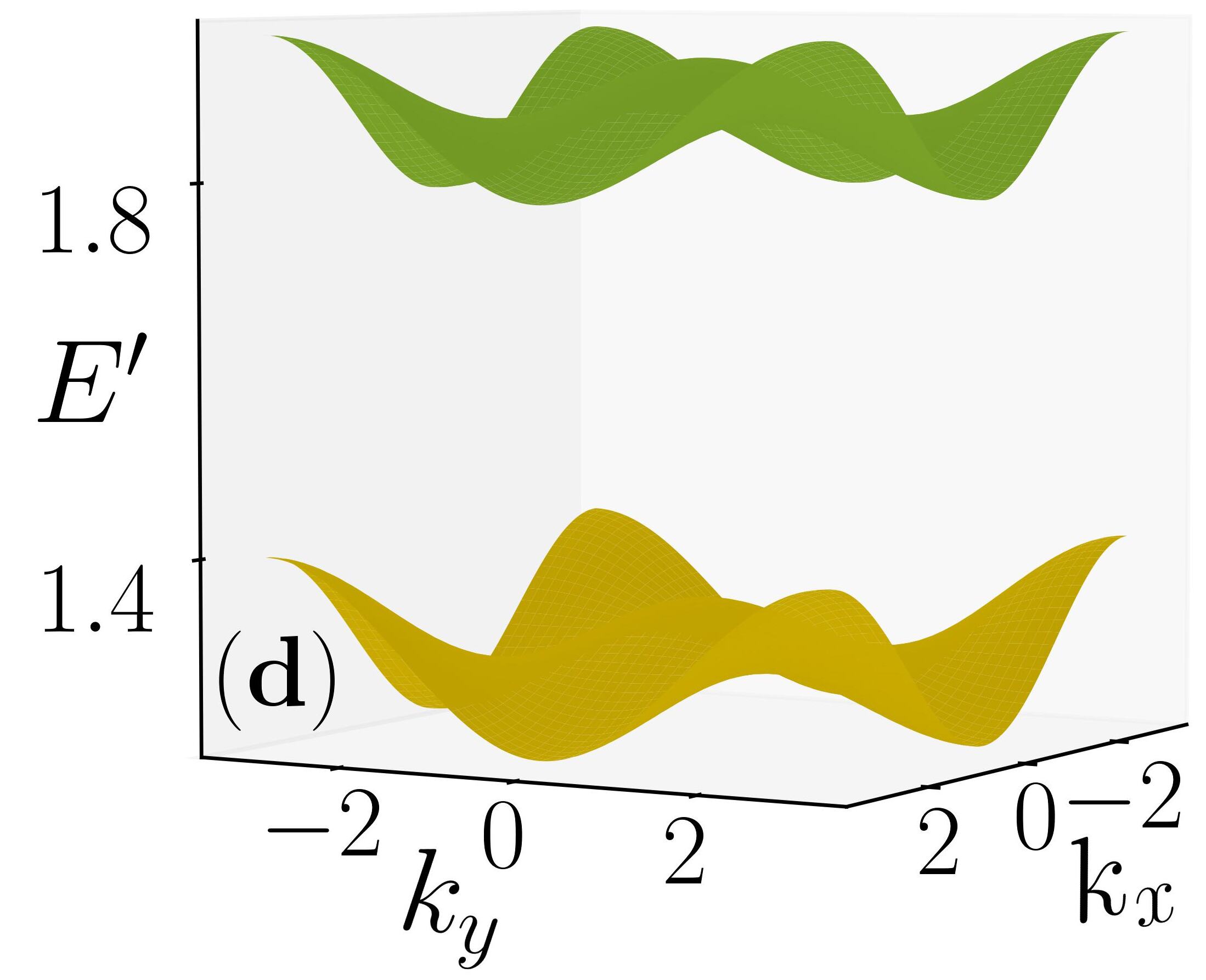}
         \captionlistentry{}
     \end{subfigure}     
    \caption{Band dispersion for one-magnon and two-magnon bound states for antiferromagnetic square-lattice. (a) Without DMI, the bands for two-magnon-bound states touch inside the BZ. (b) After applying finite DMI, with $D/J_1^z = 0.1$, the lower three bands are completely separated from each other. However, the upper two bands touch each other at the corner of the BZ. Band dispersion for two one-magnon bands for (c) $B_0=0$ and (d) $B_0/J_1^z=0.3$ have been shown at $D/J_1^z = 0.1$. Values for other parameters follow $B_0/|J_1^z|=0.3$, $J_2^z/J_1^z=0.2$, $J_1/J_1^z=0.1$, $J_2/J_1^z=0.05$. Band energy is scaled here, with $E'=E/|J_1^z|$. }
    \label{dispersion_antiferro}
\end{figure}

 As we have seen in a ferromagnet, in antiferromagnets, the presence of 
DMI breaks spin conservation and is responsible for the hybridization between one- and two-magnon bound states. However, the effective TRS is broken in presence of an external magnetic field, leading to a topological phase transition. The main difference with the ferromagnetic case is, due to a staggered spin orientation, one unit cell now comprises two different
sublattices, namely, A ($\uparrow$-spin) and B ($\downarrow$-spin). Moreover, because of this structure, there are four non-equivalent neighboring sites of B around any of A sublattice sites. Hence, we must deal with two one-magnon bands instead of one, along with four two-magnon bound states instead of two (see details in Appendix~\ref{appendix_antiferro}).
Without DMI, both the pairs of bands, at the bottom and at the top touch each at the edge of BZ (see Fig.~\ref{dispersion_antiferro}(a)), while the two intermediate ones also touch somewhere inside the BZ. Further, from Figs.~\ref{dispersion_antiferro}(c) and \ref{dispersion_antiferro}(d), we notice that the two one-magnon spectra are gapless without an external magnetic field which however become gapped in its presence.

Now, in presence of a finite DMI, the four bands are separated from each other. In Fig.~\ref{dispersion_antiferro}(b), for isotropic DMI ($\alpha = 1$) with $D/J_1^z = 0.1$, when observed from the bottom of the spectra, the first, second, and the third bands are completely separated from each other. However, the third and the fourth bands touch each other at the corner of the BZ. In the presence of an
anisotropic DMI ($\alpha \neq 1$), these four bands exhibit a variety of gap-closing and gap-opening scenarios. For concreteness, we give the following example. For $\alpha = 1$, the top two bands touch each other which get gapped for a finite $\alpha$. However, at the same anisotropy the two intermediate bands touch each other. At larger anisotropies, the same occurs for other pairs of bands.

\begin{figure}[t]
\centering
    \includegraphics[width=1\columnwidth]{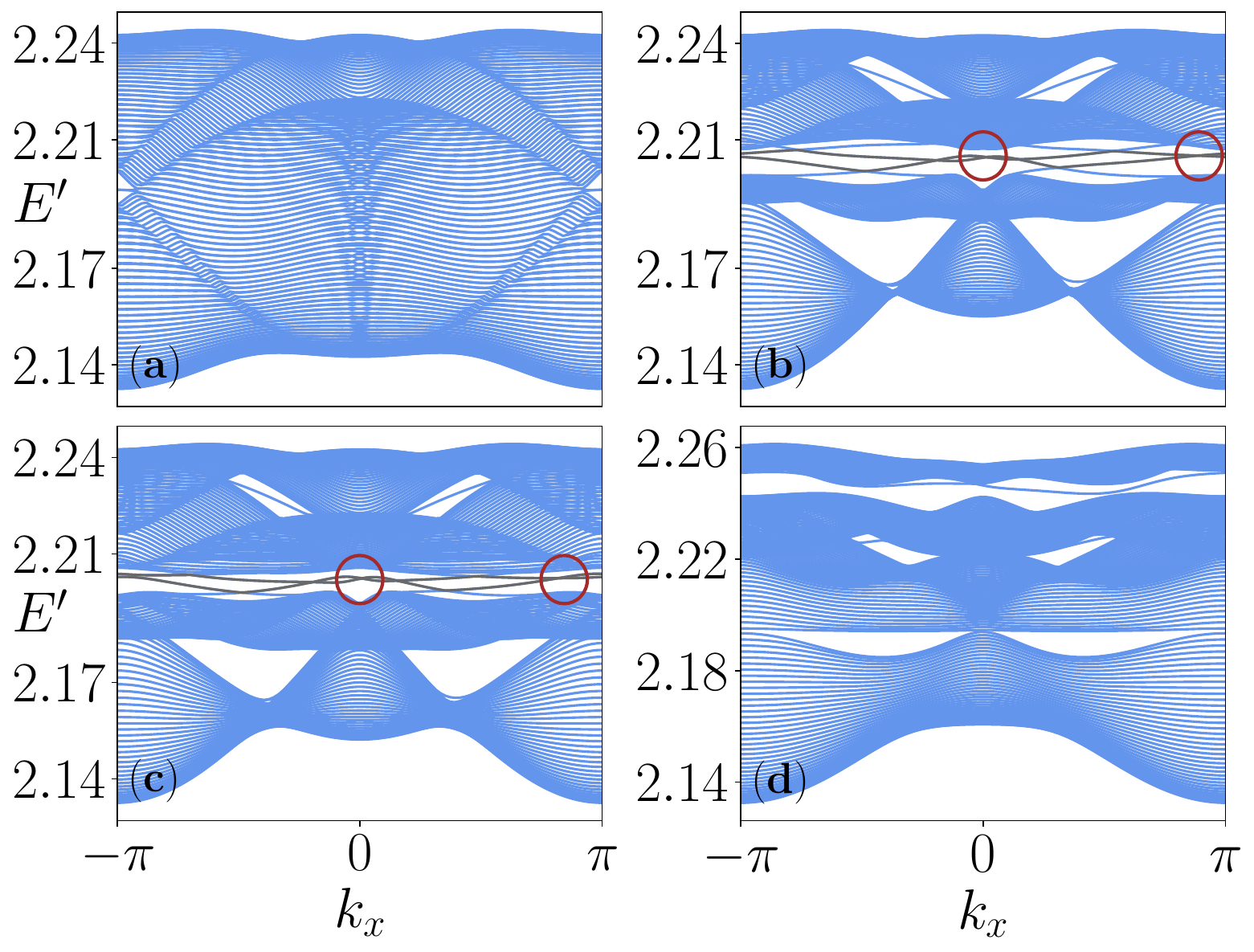}
     \caption{Two-magnon band dispersion for antiferromagnetic square lattice for semi-infinite ribbon system. (a) Without DMI, we get bands touching each other and no presence of chiral edge states. At $D/J_1^z=0.1$ (b) in the presence of isotropic DMI with $\alpha=1$ and (c) anisotropic DMI with $\alpha=0.8$, we can observe chiral edge states. (d) However, for $\alpha=2$, no chiral edge state was found between the two-magnon bound states.}
    \label{ribbon_antiferro}
\end{figure}

To ascertain the topological properties associated with the spectral gaps, or to confirm the existence of chiral edge states, we consider a semi-infinite ribbon which indeed yields a pair of chiral edge modes traversing between the two-magnon bound states in presence of DMI. To have a comprehensive view, we zoom into the two-magnon bound states in Fig.~\ref{ribbon_antiferro} and leave aside the two one-magnon bands. Here, we observe that the gap-closing transition occurs
at considerably low values of $\alpha$ (which we may call as $\alpha_c$) than that in the ferromagnetic case for same values of the parameters such as, $B_0/|J_1^z|$ and $D/|J_1^z|$. In particular, the presence of an anisotropy ($\alpha=0.8$), yields chiral edge states as seen in Fig.~\ref{ribbon_antiferro}(c). Whereas, for $\alpha = 2$, the chiral edge states disappear. This implies that there must have been a gap-closing transition occurring between $\alpha=1$ and $\alpha = 2$.
\begin{figure}[b]
     \includegraphics[width=1\columnwidth]{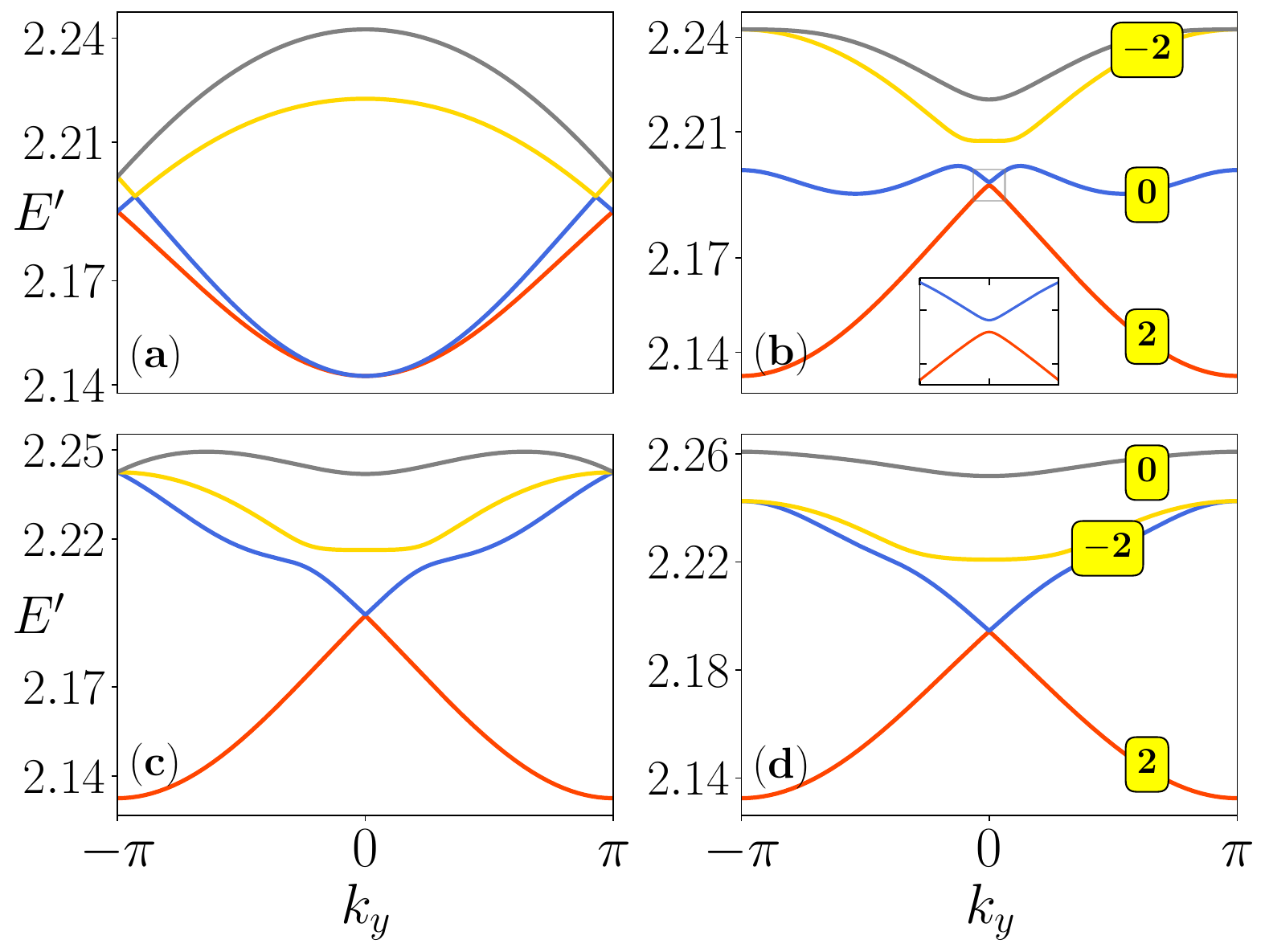}
     \caption{Band dispersion ($E'$) for four two-magnon bound states for the antiferromagnetic square lattice. (a) In the absence of DMI, all the bands touch each other along $k_y$ for $k_x = 0$. (b) In the presence of isotropic DMI, with $D/J_1^z=0.1$ and $\alpha=1$, the upper two bands touch at the corner (at $k_x=\pi$). (c) At $\alpha=1.74$, the upper three bands touch at the corner of the BZ, and (d) at $\alpha=2$, the upper band is separated from the other two bands. The yellow labels denote the Chern number of the corresponding bands.}
     \label{gapclossing_anti}
\end{figure}
Specifically, we checked the situation at an intermediate
value, namely, $\alpha = 1.74$, where we observe this spectral gap-closing, as shown in Fig.~\ref{gapclossing_anti}. To gain further clarity on the topological phases, we also calculate the Chern number separately for each of the four bands corresponding to two-magnon bound states. In the absence of DMI, the Chern numbers corresponding to all the four bands are zero. Also, for an isotropic DMI and with a small value of $D$ ($D/J_1^z = 0.1$), the Chern numbers of these bands (from bottom to top in Fig.~\ref{gapclossing_anti}(b)) are $2$, $0$, $-2$ ($-2$ is the combined Chern number of the upper two bands), respectively along with two effective crossings of chiral edge states traversing between the second and the third bands. Moreover, at the critical value of anisotropy, namely, $\alpha = \alpha_c = 1.74$,
the upper three bands touch each other at the corner of the BZ as shown in Fig.~\ref{gapclossing_anti}(c). Besides, at $\alpha = 2$, the upper band gets split from the other two. Now, the Chern numbers corresponding to the bands from bottom to top of Fig.~\ref{gapclossing_anti}(d) are $2$, $-2$ ($-2$ is the combined Chern number of two intermediate bands), $0$, respectively. The Chern number of the upper band is zero, and when we sum up the Chern numbers for the other three bands, it vanishes as well. This supports the absence of the chiral edge states between the gapped bands in Fig.~\ref{ribbon_antiferro}(d) at $\alpha=2$. Despite having non-zero Berry curvature, the Chern numbers for the one-magnon bands remain zero at each value of $\alpha$. The presence of external magnetic field $B_0$ in the Hamiltonian breaks effective TRS, which however yields finite Berry curvature beyond the gap-closing transition, and it is reminiscent of the ferromagnetic scenario.

\begin{figure}[t]
\centering
         \includegraphics[width=0.98\columnwidth]{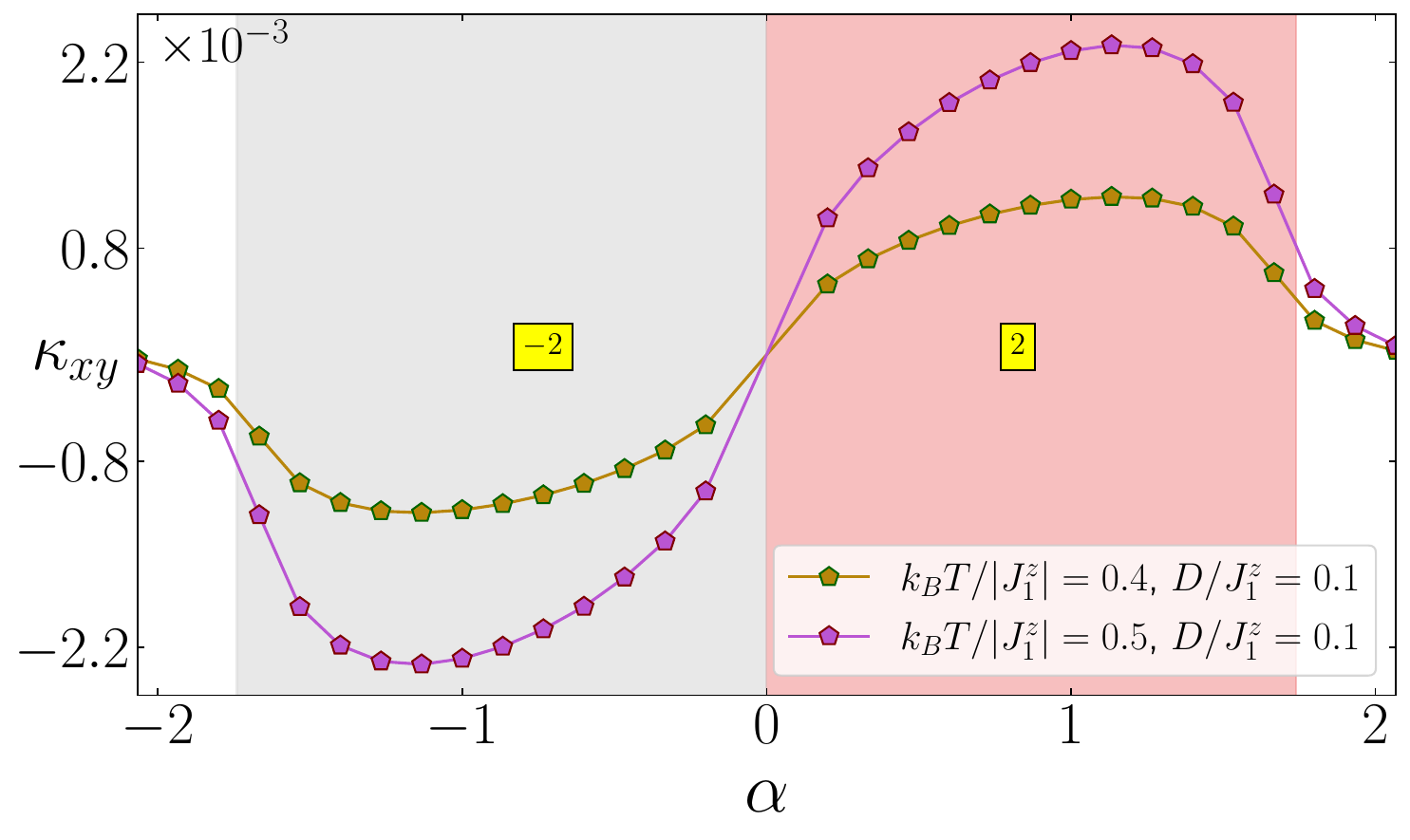}
         \caption{The behaviour of the thermal Hall conductivity as a function
            of $\alpha$ in antiferromagnetic case. The labels 2 and -2 denote the region with Chern number  $\pm 2$ of the lower band of two-magnon bound states}
         \label{Thermal_anti}
\end{figure}
Furthermore, the predominant contribution to the thermal Hall conductivity in Eq.~\eqref{kappa} comes from the lower bands of the two-magnon bound states. For two distinct temperatures ($k_BT/|J_1^z|=0.4$ and $0.5$) with $D/J_1^z= 0.1$, we observe that the thermal Hall conductivity is positive in the pink region and negative in the grey region (Fig.~\ref{Thermal_anti}). This sign change is indicative of a phase transition.
 Also, the magnitude of the thermal Hall conductivity increases with an increase in temperature. A phase transition occurs approximately at $\alpha \simeq 1.74$, which is what we have predicted earlier. However, in contrast to the ferromagnetic case, the behaviour of the Hall conductivity is somewhat different, where we observe finite thermal Hall conductivity ($\kappa_{xy}$) over a narrower region in the AFM case. Additionally, the Hall conductivity suggests different topological phases with distinct Chern numbers (here, $C = \pm 2$) relative to the ferromagnetic case ($C=\mp1$). The critical point, of course depends upon the values of other parameters used in our calculations.

\begin{figure}[t]
         \includegraphics[width=1\columnwidth]{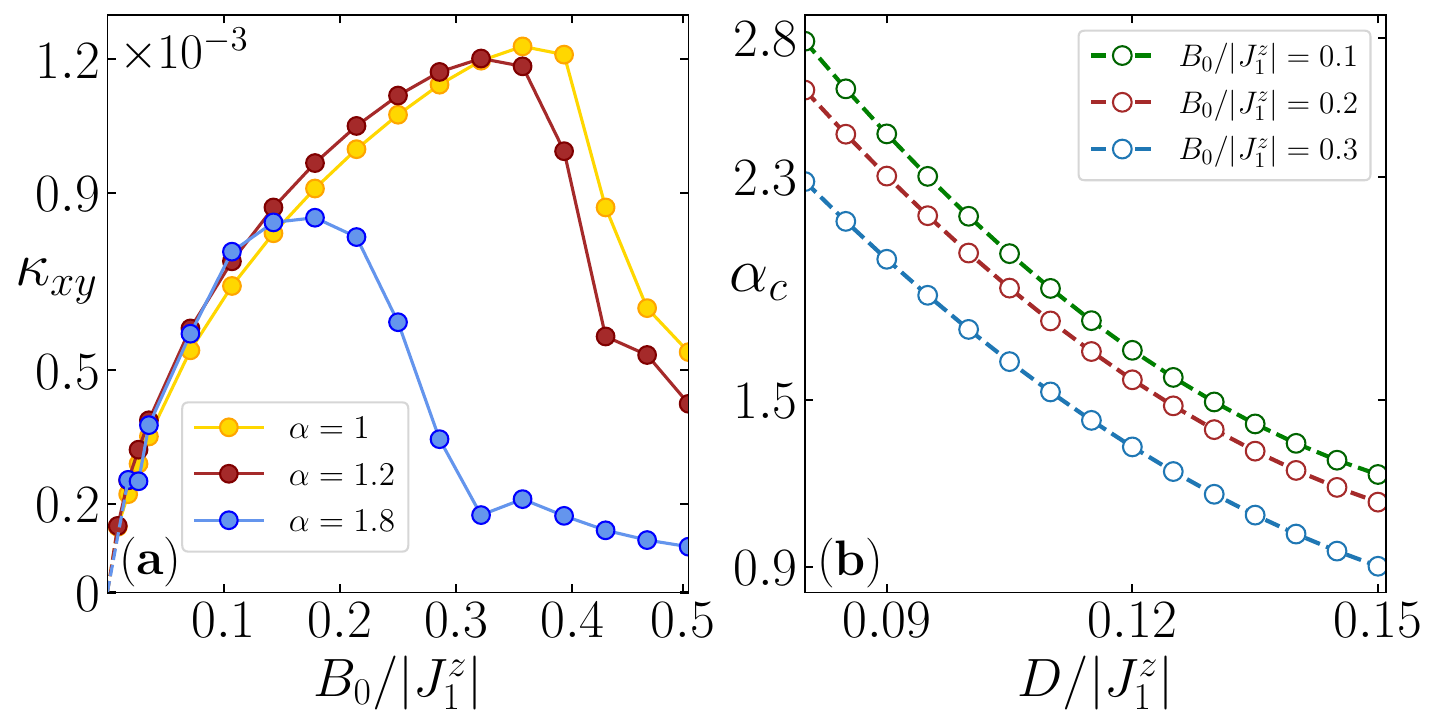}
         \caption{In antiferromagnetic case, (a) behaviour of thermal Hall conductivity as a function of $B_0$ at $k_BT=0.4$, and (b) dependency of $\alpha_c$ with $D$, for different values of $B_0$.}
         \label{kappa_vs_B_anti}
\end{figure}

Moreover, since tuning of the parameter $\alpha$ is not straightforward in experiments, variation of $\kappa_{xy}$ on the external magnetic field is a more realistic study. 
Fig.~\ref{kappa_vs_B_anti}(a) illustrates the non-monotonic variation of $\kappa_{xy}$ as a function of $B_0$. At low fields, $\kappa_{xy}$ increases with $B_0$, and beyond a certain value, it starts to decrease and finally saturates, which is clearer for the case of (as compared to lower $\alpha$ values) $\alpha= 1.8$ , just beyond the gap-closing transition point in Fig.~\ref{kappa_vs_B_anti}(a). The cases for $\alpha=1$ and $\alpha = 1.2$ will saturate at much larger values of $B_0$ (not shown here).  Notably, without an external magnetic field, the two one-magnon bands (corresponding to the A and B sublattices) are degenerate. According to the magnetic space group symmetry, at zero magnetic field, the antiferromagnetic system possesses effective TRS, which implies that the thermal Hall conductivity should be zero, even in the presence of a non-zero DMI.  

Now, to identify the topological and trivial regimes, we need to examine Fig.~\ref{kappa_vs_B_anti}(b), where we observe that for a fixed value of $D$, $\alpha_c$ (the critical value at which the phase transition occurs) decreases as the magnetic field $B_0$ increases. Moreover, for different values of $B_0$, the plot shows that $\alpha_c$ decreases with increasing the DMI ($D$) depending on the values of $B_0$, as the other exchange interaction parameters are kept fixed.


\section{CONCLUSION}
\label{conclusion}
We have shown that nontrivial band topology arises in our magnon Hamiltonian from the hybridization of bands pertaining to one- and two-magnon sectors. Our motivation was to ascertain the effects of an anisotropic DMI in both the ferromagnetic and the antiferromagnetic models for studying topological magnons. Although we have limited our study in 2D square lattice model, the formalism is readily applicable to any other 2D lattices. The band structures and in some cases a phase plot reveal that there are different topological phases, and hence, multiple transitions from one phase to another exist. Even though the DMI term is responsible for opening of a spectral gap, the nature of the gap has to be ascertained, which is done via examining the chiral edge states in ribbon geometries. We also show that the external magnetic field plays an important role in controlling the topology of the magnon bands. In presence of larger anisotropy, the system may loose its ordered ground state, leading to a trivial spectral gap. Nevertheless, our entire formalism relies on the assumption that the DMI remains sufficiently small to act as a perturbation and not to destroy the ordered ground state.

Let us enumerate the highlights for the ferromagnetic case and hence compare them with those for the antiferromagnets. In spite of the effective TRS broken via the DMI term, the ferromagnetic case demonstrates extended trivial region characterized by zero Chern number, and hence yields vanishing thermal Hall conductivity. Second, there are evidence of multiple topological phase transitions, both from topological to topological and topological to trivial phases at least in the FM case. Both are accompanied by a bulk gap-closing transition, and the latter is characterized by vanishing of the chiral edge modes at certain critical value of the anisotropic DMI. The critical anisotropy ($\alpha_c$) can be controlled via the external magnetic field. Further, in the ferromagnetic case, a large magnetic field helps to maintain its ordered spin state and prevent the formation of spin spirals or skyrmions. So, for a higher magnetic field, $\alpha_c$ becomes larger, and the gap-closing transition occurs at a higher value of $\alpha_c$. On the contrary, despite a similar scenario with gap-opening and gap-closing transitions, in the antiferromagnetic case, all the characteristics depend on a higher number of two-magnon bands with non-zero Chern numbers. Therefore, drawing a phase plot solely based on the Chern number of a single band would be insufficient in this case. Nonetheless, for smaller values of the anisotropy parameter $\alpha$, we observe the emergence of two chiral edge states between two fully gapped bands. Furthermore, for larger values of the magnetic field, topological to trivial phase transitions occur easily, justifying $\alpha_c$ to have an inverse proportionality with $B_0$.

At a fundamental level, our results focus on the topological behaviour of low-energy spin excitations in a magnetically ordered system, and the effects of anisotropic DMI and a magnetic field are ascertained on the behaviour of the thermal Hall effect. While the formalism was introduced in Ref.~\cite{magnonboundpairs2023} for the ferromagnet, the corresponding scenario in AFM is much more involved. Though topological phase transitions obtained in both are qualitatively similar, our work extends the calculations for AFM case.

\appendix 
\section{Introducing three- and four-magnon states}
\label{appendix:three- and four-magnon}
We have described the notation associated with the one- and two-magnon states earlier. Now for a ferromagnet, $\Delta S = 3$ and $\Delta S = 4$ sectors are given by,
$\ket{i, j, k} = (1/\sqrt{2S})^3 S_i^-S_j^-S_k^-\ket{0}$ and $\ket{i,j,k,l}=(1/\sqrt{2S})^4 S_i^-S_j^-S_k^-S_l^-\ket{0}$, which may be referred to as three- and four-magnon excitations, respectively. However, in the AFM system, there are distinct ways to define three- and four-magnon states. At three-magnon excitation can be produced via four different ways, such as,
\begin{equation}
|i, j, k\rangle =
\begin{cases}
\left(\frac{1}{\sqrt{2 S}}\right)^3 S_i^{-(+)} S_j^{-(+)} S_k^{-(+)} \ket{0},\\
\left(\frac{1}{\sqrt{2 S}}\right)^3 S_i^{-(+)} S_j^{+(-)} S_k^{-(+)} \ket{0},
\end{cases}
\end{equation}
where all the lattice sites belong to either A or B sublattice in the first case, and any two sites belong to same sublattice for the second case. Further, the four-magnon states also can be created via different types of spin-flips, namely,
\begin{equation}
|i, j, k, l\rangle =
\begin{cases}
\left(\frac{1}{\sqrt{2 S}}\right)^4 S_i^{-(+)} S_j^{-(+)} S_k^{-(+)} S_l^{-(+)}\ket{0},\\
\left(\frac{1}{\sqrt{2 S}}\right)^4 S_i^{-(+)} S_j^{-(+)} S_k^{-(+)} S_l^{+(-)} \ket{0},\\
\left(\frac{1}{\sqrt{2 S}}\right)^4 S_i^{-(+)} S_j^{-(+)} S_k^{+(-)} S_l^{+(-)} \ket{0},
\end{cases}
\end{equation}
which suggests the types of events: either all the states are from the same sublattice, any three of them are from the same sublattice or any two of them are from the same sublattice sites (A or B).

\section{Derivation of Effective Hamiltonian}
\label{appendix:A}

In the case of strong spin anisotropy, where we have taken a large value for neighbouring Heisenberg exchange interaction, the one-magnon and two-magnon bound states, separated from the continuum, allow us to write an effective Hamiltonian, with

\begin{align}
    H = H_0 +V,\label{H}
\end{align}
where,
\begin{equation}
\begin{split}
    H_0 = \sum_{p=1}^{N_x}\sum_{q = 1}^{N_y} \left( J_1^z S_{p,q}^z S_{p+1, q}^z\right. &+ \left.J_1^z S_{p,q}^z S_{p, q+1}^z\right.\\
    &\left.-B_0 S_{p,q}^z \right) 
\end{split}
\end{equation}
and,
 \begingroup
 \allowdisplaybreaks
\begin{align}
    &V = \sum_{p=1}^{N_x}\sum_{q = 1}^{N_y} \left[ J_1 ( S_{p,q}^x S_{p+1,q}^x + S_{p,q}^y S_{p+1,q}^y + S_{p,q}^x S_{p,q+1}^x +\right.\nonumber \\  
    &\left.S_{p,q}^y S_{p,q+1}^y\right) + \mb{S}_{p,q}\cdot \mb{I}_2 (\mb{S}_{{p+2},q} + \mb{S}_{p,q+2}) \nonumber \\ 
    & + \alpha D \hat{y} \cdot \left( \mb{S}_{p,q} \times \mb{S}_{p+1,q} \right) + D \hat{x} \cdot \left( \mb{S}_{p,q} \times \mb{S}_{p,q+1} \right)]. 
\end{align}
\endgroup
$H_0$ is responsible for the formation of magnon bounds pairs and only includes the $z$-component of the nearest neighbour Heisenberg exchange interaction. Other spin-spin interactions, along with the DMI term, are included in $V$, and it is treated perturbatively.~The matrix elements of effective low-energy Hamiltonian can be represented as~\cite{cohen1998atom},
\begin{align}
     &H_{\text{eff}}^{\mu \nu} = \bra{\mu}  H_0\ket{\nu} + \bra{\mu}  V \ket{\nu}\nonumber\\
     &+ \frac{1}{2} \sum_\nu \bra{\mu} V \ket{\gamma} \bra{\gamma} V \ket{\nu} \left( \frac{1}{E_\mu - E_\gamma} + \frac{1}{E_\nu - E_\gamma}\right).\label{heff}
\end{align}
Here, $\ket{\mu}$ and $\ket{\nu}$ are the eigenstates of $H_0$ with eigenvalues $E_\mu = \bra{\mu}H_0\ket{\mu}$ and $E_\nu = \bra{\nu}H_0\ket{\nu}$ respectively, which correspond to the single-magnon excitation energy and two-magnon bound state energy. As we are interested in a low energy Hamiltonian,~$\ket{\gamma}$ denotes the intermediate states with energy $E_\gamma = \bra{\gamma}H_0\ket{\gamma}$, which reside in the two-magnon continuum.

However, in the AFM case, the intermediate state $\ket{\gamma}$ (see Eq.~\eqref{heff}) in addition to the two-magnon continuum, possesses the low-energy three-magnon and four-magnon states as well, whose energies are lower than those of the individual magnons in presence of NN Heisenberg interaction. The three-magnon states are indispensable for determining the hopping energies between one-magnon states on the same sublattice. Further, the low-energy four-magnon states are responsible for the hopping between two two-magnon bound states. Notably, the energy of a four-magnon intermediate state should be less than the combined energy of two two-magnon bound states.


\subsection{Hopping Model in a Ferromagnet }
\label{appendix_ferro}
\begin{figure}[b]
         \includegraphics[width=0.9\columnwidth]{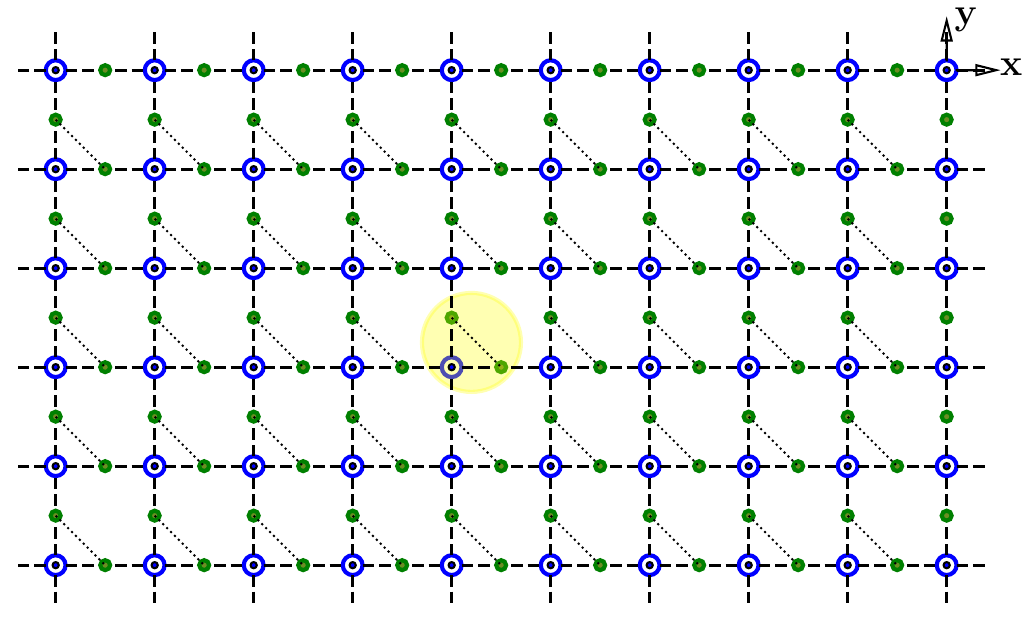}
         \caption{Effective hopping model for a ferromagnetic square lattice model}
         \label{Ferromagnet1}
\end{figure}

In the case of a ferromagnetic,
we shall introduce a vector, $c^\dg_{i,j} = (s^\dg_{i,j}, x^\dg_{i,j}, y^\dg_{i,j})$, which comprises of single spin-flip $s^\dg_{i,j}$,
and double spin-flip particle creators,
$x^\dg_{i,j}$ and $y^\dg_{i,j}$. If $\ket{0}$ is defined as the initial ground state, then those three particle creators can be defined as $\ket{i,j} = s^\dg_{i,j}\ket{0}$, $\ket{i,j;i+1,j} = x^\dg_{i,j}\ket{0}$, and $\ket{i,j;i,j+1} = y^\dg_{i,j}\ket{0}$ respectively.
In Fig.~\ref{Ferromagnet1}, we illustrate the ferromagnetic hopping model schematically~\cite{magnonboundpairs2023}. The blue dots represent the original spin locations, which are also the expected positions for single spin-flip (one-magnon excitation), while the green dots indicate positions of the double spin-flips (two-magnon excitations). The yellow highlighted region denotes the basis vectors within a single unit cell.
Evaluating Eq.~(\ref{heff}) for all possible values of $\ket{\mu}$, $\ket{\nu}$, and $\ket{\gamma}$, we can reduce the Hamiltonian in Eq.~(\ref{H}) to an effective hopping model~\cite{magnonboundpairs2023}, given by,
 \begingroup
 \allowdisplaybreaks
\begin{align}
    &H_{\text{\tiny{eff}}}^{\text{\tiny{F}}} = \sum\limits_{i,j}[\mb{c}^\dagger_{i,j} \mb{m}\hspace{1pt}\mb{c}_{i,j} + \left(\mb{c}^\dagger_{i+1,j} \mb{t}_{x_1} \mb{c}_{i,j} + \text{H.c.}\right)\nonumber\\
    &+ \left(\mb{c}^\dagger_{i,j+1} \mb{t}_{y_1} \mb{c}_{i,j} + \text{H.c.}\right) + \left(\mb{c}^\dagger_{i+2,j} \mb{t}_{x_2} \mb{c}_{i,j} + \text{H.c.}\right)\nonumber\\
    &+ \left(\mb{c}^\dagger_{i,j+2} \mb{t}_{y_2} \mb{c}_{i,j} + \text{H.c.}\right) + \left(\mb{c}^\dagger_{i+1,j-1} \mb{t}_{x\bar{y}}\mb{c}_{i,j} + \text{H.c.}\right)]. \label{hopping_H_eff_ferro}
\end{align}
\endgroup

Here, the spin-polarized state may be defined as the $\ket{0}$ state. Now for spin-$\frac{1}{2}$ particle the single spin-flip will be denoted as, $S^{-}_{i,j}\ket{0} = \ket{i,j}$, known as single-magnon and double spin-flips will be denoted by $S^{-}_{i,j}S^{-}_{l,m}\ket{0} = \ket{i,j;l,m}$, referred to the two-magnon state. Accordingly, we shall obtain the hopping matrices corresponding to the Eq.~(\ref{hopping_H_eff_ferro}). Note that one can recover results for Ref.~\cite{magnonboundpairs2023} corresponding to $\alpha=1$. The matrices are,
 \begin{subequations}
 \begingroup
 \allowdisplaybreaks
\begin{align}
\small
    \mathbf{m}&=\begin{pmatrix}
                    B_0-2(J_1^z + J_2^z) & -\frac{\alpha D}{2} & -i \frac{D}{2}\\
                    -\frac{\alpha D}{2} & u & \frac{J_1^2}{2 J_1^z}\\
                    i \frac{D}{2} & \frac{J_1^2}{2 J_1^z} & u
                \end{pmatrix},\\
    \mathbf{t}_{x_1}&=\begin{pmatrix}
                        \frac{J_1}{2} & \frac{\alpha D}{2} & 0\\
                        0 & \frac{J_2}{2} + \frac{J_1^2}{4J_1^z} & 0\\
                        0 & \frac{J_1^2}{2 J_1^z} & \frac{J_1^2}{2 J_1^z}      
                  \end{pmatrix},
    \mathbf{t}_{x_2}=\begin{pmatrix}
                        \frac{J_2}{2} & 0 & 0\\
                        0 & \frac{J_2^2}{4J_1^z} & 0\\
                        0 & 0 & \frac{J_2^2}{2J_1^z}
                 \end{pmatrix},\\              
    \mathbf{t}_{y_1}&=\begin{pmatrix}
                        \frac{J_1}{2} & 0 & i\frac{D}{2}\\
                        0 & \frac{J_1^2}{2J_1^z} & \frac{J_1^2}{2J_1^z}\\
                        0 & 0 & \frac{J_2}{2} + \frac{J_1^2}{4J_1^z}
                 \end{pmatrix},
    \mathbf{t}_{y_2}=\begin{pmatrix}
                        \frac{J_2}{2} & 0 & 0\\
                        0 & \frac{J_2^2}{2J_1^z} & 0\\
                        0 & 0 & \frac{J_2^2}{4J_1^z}
                 \end{pmatrix},\\  
    \mathbf{t}_{x\bar{y}}&=\begin{pmatrix}
                        0 & 0 & 0\\
                        0 & 0 & 0\\
                        0 & \frac{J_1^2}{2J_1^z} & 0
                 \end{pmatrix},                               
\end{align}
\label{hopping_ferro}
\endgroup
\end{subequations}
where,
\begin{equation}
\small
    u = 2B_0 -3J_1^z + \frac{3J_1^2}{2J_1^z} -4J_2^z + \frac{3J_2^2}{2J_1^z}.
\end{equation}   

For an infinite square lattice under periodic boundary conditions, we apply Fourier transformation,
\begin{subequations}
 \begingroup
 \allowdisplaybreaks
    \begin{align}
        \mathbf{c}_{i,j} = \frac{1}{\sqrt{N}} \sum_\mathbf{k} e^{i\mathbf{k}\cdot \mathbf{r}_{i,j}}\mathbf{c}_{\mathbf{k}},\\
        \mb{c}^\dg_{i,j} = \frac{1}{\sqrt{N}} \sum_\mathbf{k} e^{i\mathbf{k}\cdot \mathbf{r}_{i,j}}\mathbf{c}_{\mathbf{k}}.
    \end{align}
    \endgroup
\end{subequations}
The effective Hamiltonian in reciprocal space looks like,  
\begin{equation}
\begin{split}
    H_{\text{\tiny{F}}}'(\mathbf{k}) = & \mathbf{m} + e^{-ik_x}\mathbf{t}_{x_1} + e^{-ik_y}\mathbf{t}_{y_1} \\
    & + e^{-i2k_x}\mathbf{t}_{x_2} + e^{-i2k_y}\mathbf{t}_{y_2} + e^{-i(k_x-k_y)}t_{x\bar{y}} + \text{H.c}.
\end{split}
\label{R}
\end{equation}

Simplification of Eq.~\eqref{R} obtained via a unitary transformation, $H_{\text{\tiny{F}}}(\mb{k})=\mathbf{U}^\dg H_{\text{\tiny{F}}}'(\mb{k})\mathbf{U}$, where, $\mathbf{U}=\text{diag}(1, e^{\frac{ik_x}{2}}, e^{\frac{ik_y}{2}})$ yields the effective Hamiltonian for ferromagnet (Eq.~\eqref{effective_ferro}).

\subsection{Hopping Model in an Antiferromagnet}
\label{appendix_antiferro}
In case of an antiferromagnet, the staggered orientation of spins at neighbouring lattice sites is defined as the $\ket{0}$ state and at an arbitrary $(i, j)^{\text{th}}$ location, the spin orientation could either be $\uparrow$ or $\downarrow$ depending on whether it belongs to an A or a B sublattice. Hence, distinct from the ferromagnetic case, we now have four non-equivalent B sublattice sites around each A sublattice. Such a scenario yields four two-magnon bound states and two one-magnon excitations in a unit cell.

Let us assume that there is a $\uparrow$-spin (A sublattice) in the $(i, j)^{\text{th}}$ location. An identical result follows if it is a B sublattice site hosting a $\downarrow$-spin. One-magnon excitations will occur at two distinct positions, namely, the $(i, j)^{\text{th}}$ and the $(i+1, j)^{\text{th}}$ sites. We also expect one-magnon excitations to manifest at three other non-equivalent positions in the unit cell, and all of these will be energetically degenerate. Therefore, we introduce a basis $c_{ij}^{\dagger} = \lbrace s_{ij}^\dagger, s_{i+1,j}^\dagger, x_{ij}^\dagger(+), x_{i,j}^\dagger(-), y_{ij}^\dagger(+), y_{ij}^\dagger(-)\rbrace$, where $s_{ij}^\dagger$ and $s_{i+1,j}^\dagger$ are the basis vectors corresponding to a single-magnon excitation and are denoted by,
\begin{equation}
\begin{split}
    \ket{i,j} &= s_{i,j}^\dagger\ket{0}=S_{i,j}^-(\uparrow)\ket{0}\\
    \ket{i+1,j} &= s_{i+1,j}^\dagger\ket{0}=S_{i+1,j}^+(\downarrow)\ket{0}.
\end{split}
\end{equation}
  The $\uparrow$ and $\downarrow$ spins inside the brackets denote the ground state spin orientation at that location. Whereas the other four basis vectors $x_{ij}^\dagger(+), x_{i,j}^\dagger(-), y_{ij}^\dagger(+)$ and $ y_{ij}^\dagger(-)$ correspond to the two-magnon excitations and are respectively denoted by,
  \begin{subequations}
  \begingroup
  \allowdisplaybreaks
  \begin{align}
      \ket{i,j;i+1,j} &= S_{ij}^-(\uparrow)S_{i+1,j}^+(\downarrow)\ket{0},\\
      \ket{i,j;i-1,j} &= S_{ij}^-(\uparrow)S_{i-1,j}^+(\downarrow)\ket{0},\\
      \ket{i,j;i,j+1} &= S_{ij}^-(\uparrow)S_{i,j+1}^+(\downarrow)\ket{0},\\
      \ket{i,j;i,j-1} &= S_{ij}^-(\uparrow)S_{i,j-1}^+(\downarrow)\ket{0}.
      \end{align}
      \endgroup
  \end{subequations}

  \begin{figure}[b]
         \includegraphics[width=0.9\columnwidth]{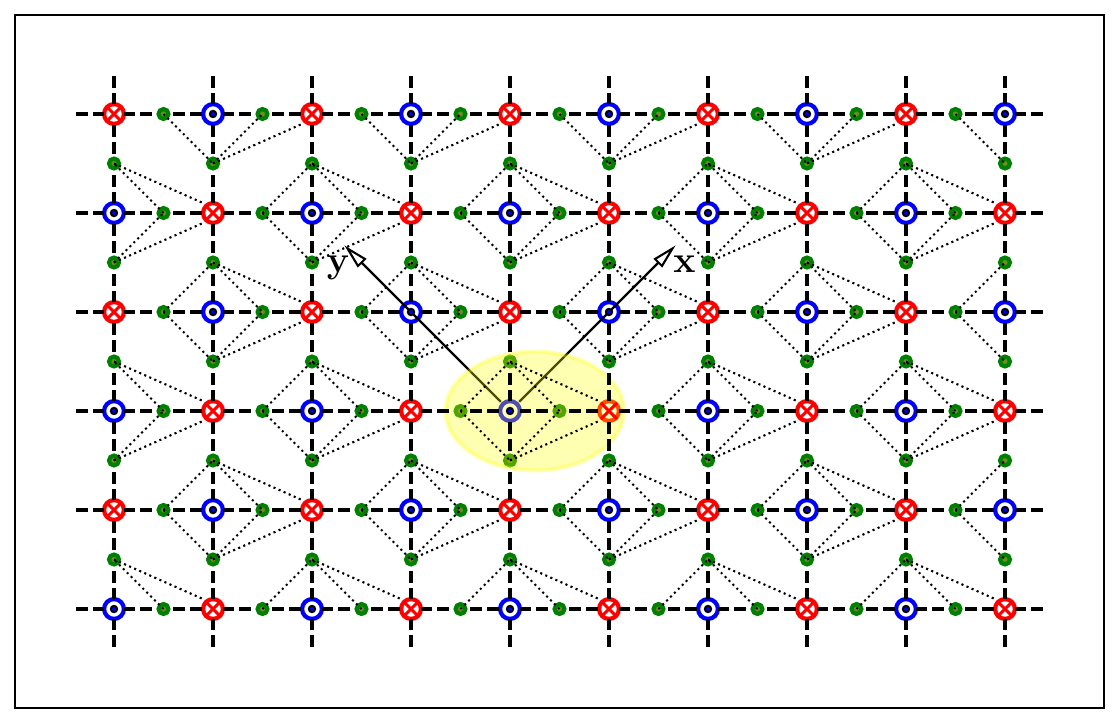}
         \caption{Effective Hopping Model for an Antiferromagnetic square lattice model}
         \label{AntiFerromagnet}
\end{figure}
In Fig.~\ref{AntiFerromagnet}, we illustrate the antiferromagnetic hopping model schematically. The blue dots and the red dots represent the spin-$\uparrow$ (A) and spin-$\downarrow$ (B) locations, respectively, while the green dots indicate positions of the double spin-flips at the neighboring sites. The yellow highlighted region denotes the basis vectors within a single unit cell of this effective hopping model. We can reduce the Hamiltonian $H = H_0 +V$ to an effective hopping model for the antiferromagnetic case, with the non-zero hopping matrices, given by,
\begin{align}
    &H_{\text{\tiny{eff}}}^{\text{\tiny{AF}}} = \sum\limits_{i,j}[\mb{c}^\dagger_{i,j} \mb{m}\hspace{1pt}\mb{c}_{i,j} + \left(\mb{c}^\dagger_{i+1,j} \mb{t}_{x_1} \mb{c}_{i,j} + \text{H.c.}\right)\nonumber\\
    &+\left(\mb{c}^\dagger_{i,j+1} \mb{t}_{y_1} \mb{c}_{i,j} + \text{H.c.}\right) + \left(\mb{c}^\dagger_{i+1,j-1} \mb{t}_{x\bar{y}} \mb{c}_{i,j} + \text{H.c.}\right)\nonumber\\
    & \quad \quad +\left(\mb{c}^\dagger_{i+1,j+1}\right. \mb{t}_{xy} \mb{c}_{i,j} + \left. \text{H.c.}\right).
    \label{hopping_H_eff_antiferro}
\end{align}
    Here, we get the hopping matrices as,
\begin{widetext}
\begin{subequations}

\begingroup
\allowdisplaybreaks
    \begin{align}
    \mathbf{m}&=\begin{pmatrix}
                    2J_1^z -2J_2^z +B_0 & 0 & -\frac{D_x}{2} & \frac{D_x}{2} & i\frac{D_y}{2} & -i\frac{D_y}{2}\\
                    0 & 2J_1^z -2J_2^z -B_0 & -\frac{D_x}{2} & 0 & 0 & 0\\
                    -\frac{D_x}{2} & -\frac{D_x}{2} & u' & \frac{J_2}{2} & - \frac{J_1^2}{4J_1 ^z} & - \frac{J_1^2}{4J_1 ^z} \\
                    \frac{D_x}{2} & 0 & \frac{J_2}{2} & u' & - \frac{J_1^2}{4J_1 ^z} & - \frac{J_1^2}{4J_1 ^z}\\
                    -i\frac{D_y}{2} & 0 & - \frac{J_1^2}{4J_1 ^z} & - \frac{J_1^2}{4J_1 ^z} & u' & \frac{J_2}{2}\\
                    i\frac{D_y}{2} & 0 & - \frac{J_1^2}{4J_1 ^z} & - \frac{J_1^2}{4J_1 ^z} & \frac{J_2}{2} & u'   
                \end{pmatrix},\\           
    \mathbf{t}_{x_1}&=\begin{pmatrix}
                        -\frac{J_1^2}{2J_1^z + 2B_0} & 0 & 0 & 0 & 0 & 0\\
                        0 & -\frac{J_1^2}{2J_1^z - 2B_0} & 0 & 0 & 0 & 0\\
                        0 & 0 & 0 & 0 & 0 & 0\\
                        0 & 0 & -\frac{J_1^2}{4J_1^z} & 0 & -\frac{J_1^2}{4J_1^z} & 0\\
                        0 & 0 & 0 & 0 & 0 & 0 \\
                        0 & -i\frac{D_y}{2} & -\frac{J_1^2}{4J_1^z} & 0 & -\frac{J_1^2}{4J_1^z} & 0
                  \end{pmatrix}, \hspace{5 pt}
    \mathbf{t}_{y_1}=\begin{pmatrix}
                        -\frac{J_1^2}{2J_1^z + 2B_0} & 0 & 0 & 0 & 0 & 0\\
                        0 & -\frac{J_1^2}{2J_1^z - 2B_0} & 0 & 0 & -i\frac{D_y}{2} & 0\\
                        0 & 0 & 0 & -\frac{J_1^2}{4J_1^z} & -\frac{J_1^2}{4J_1^z} & 0\\
                        0 & 0 & 0 & 0 & 0 & 0\\
                        0 & 0 & 0 & 0 & 0 & 0\\
                        0 & 0 & 0 & -\frac{J_1^2}{4J_1^z} & -\frac{J_1^2}{4J_1^z} & 0
                 \end{pmatrix},\\                  
    & \hspace{2cm}\mathbf{t}_{x\bar{y}} =\begin{pmatrix}
                        \frac{J_2}{2} - \frac{J_1^2}{4J_1^z + 4B_0} & 0 & 0 & 0 & 0 & 0\\
                        0 & \frac{J_2}{2} - \frac{J_1^2}{4J_1^z - 4B_0}& 0 & 0 & 0 & 0\\
                        0 & 0 & -\frac{J_2^2}{4J_1^z} & 0 & 0 & 0\\
                        0 & \frac{D_x}{2} & \frac{J_2}{2} & -\frac{J_2^2}{4J_1^z} & 0 & 0\\
                        0 & 0 & 0 & 0 & -\frac{J_2^2}{2J_1^z} & 0\\
                        0 & 0 & 0 & 0 & 0 & -\frac{J_2^2}{2J_1^z}
                 \end{pmatrix},\\
    & \hspace{2cm} \mathbf{t}_{xy}=\begin{pmatrix}
                        \frac{J_2}{2} - \frac{J_1^2}{4J_1^z + 4B_0} & 0 & 0 & 0 & 0 & 0\\
                        0 & \frac{J_2}{2} - \frac{J_1^2}{4J_1^z - 4B_0} & 0 & 0 & 0 & 0\\
                        0 & 0 & -\frac{J_2^2}{2J_1^z} & 0 & 0 & 0\\
                        0 & 0 & 0 & -\frac{J_2^2}{2J_1^z} & 0 & 0\\
                        0 & 0 & 0 & 0 & -\frac{J_2^2}{4J_1^z} & 0\\
                        0 & 0 & 0 & 0 & \frac{J_2}{2} & -\frac{J_2^2}{4J_1^z}
                 \end{pmatrix}                 
    \end{align}
    \label{hopping_anti}
    \endgroup
\end{subequations}

where, 
\begin{equation}
    u'=3J_1^z -\frac{J_1^2}{2J_1^z} -4J_2^z -\frac{3J_2^2}{2J_1^z}. 
\end{equation}
The effective Hamiltonian in reciprocal space is again given by 
\begin{equation}
    H_{\text{\tiny{AF}}}'(\mathbf{k}) =  \mathbf{m} + e^{-ik_x}\mathbf{t}_{x_1} + e^{-ik_y}\mathbf{t}_{y_1} + e^{-i(k_x-k_y)}t_{x\bar{y}}                          + e^{-i(k_x+k_y)}t_{xy} + \text{H.c},
\label{R2}
\end{equation}
        and after a similar unitary transformation with
\begin{equation}
    \mathbf{U}=\text{diag}(1, e^{i\left(k_x/2 -k_y/2\right)}, e^{i\left(k_x/4 -k_y/4\right)}, e^{-i\left(k_x/4 -k_y/4\right)}, e^{i\left(k_x/4 + k_y/4\right)}, e^{-i\left(k_x/4 + k_y/4\right)}),
\end{equation}
the final Hamiltonian is given in Eq.~\eqref{effective_antiferro}.

\end{widetext}


\bibliographystyle{apsrev4-2}   
\bibliography{main}

\begin{thebibliography}{60}
\bibitem [{In our case()}]{note}%
  \BibitemOpen
  In our case,\ \href@noop {} {\bibinfo {title} {the phase transition is even
  in $\alpha$, which implies that at $\alpha=\pm 3.6$ we observe a gap closing
  scenario.}}\BibitemShut {Stop}%
\bibitem [{In order to lucidly visualize different topological phase
  transitions()}]{note1}%
  \BibitemOpen
  In order to lucidly visualize different topological phase transitions,\
  \href@noop {} {\bibinfo {title} {it is more appropriate to go the phase
  diagram defined by $\uppercase{D}_x$ and $\uppercase{D}_y$.}}\BibitemShut
  {Stop}%
\end{thebibliography}

\end{document}